\documentclass[prl,aps,twocolumn,superscriptaddress]{revtex4-2}

\usepackage{amsmath,amssymb,amsfonts,bbm,color,graphicx}

\usepackage[utf8]{inputenc}
\usepackage[T1]{fontenc}

\usepackage[unicode=true,colorlinks=true]{hyperref}
\hypersetup{linkcolor=blue,citecolor=blue,urlcolor=blue}

\usepackage{bm}
\usepackage{dsfont}
\usepackage{xcolor}
\usepackage{wasysym}

\usepackage{braket}

\renewcommand{\Re}{\operatorname{Re}}

\newcommand{\Au}{\mathrm{A}} 
\newcommand{\Bu}{\mathrm{B}} 

\newcommand{\iu}{\mathrm{i}} 
\newcommand{\eu}{\mathrm{e}} 
\newcommand{\hc}{\mathrm{h.c.}} 

\newcommand{\bvec}[1]{\bm{#1}}

\newcommand{\Jki}{J_\mathrm{K}^{(\mathrm{I})}}
\newcommand{\Jkii}{J_\mathrm{K}^{(\mathrm{II})}}
\newcommand{\ji}{\mathcal{J}_\mathrm{I}}

\begin{document}

\title{Spin-polarons and ferromagnetism in doped dilute moiré-Mott insulators}

\author{Urban F. P. Seifert}
\affiliation{Kavli Institute for Theoretical Physics, University of California, Santa Barbara, CA 93106, USA}
\author{Leon Balents}
\affiliation{Kavli Institute for Theoretical Physics, University of California, Santa Barbara, CA 93106, USA}
\affiliation{Canadian Institute for Advanced Research, Toronto, Ontario, Canada}

\begin{abstract}%
Moiré heterostructures of transition metal dichalcogenides (TMD) exhibit Mott-insulating behaviour both at half-filling as well as at fractional fillings, where electronic degrees of freedom form self-organized Wigner crystal states. An open question concerns magnetic states obtained by lifting the pseudospin-1/2 degeneracy of these states at lowest temperatures. While at half-filling virtual hopping is expected to induce (weak) antiferromagnetic exchange interactions, these are strongly suppressed when considering \emph{dilute} filling fractions. We argue that instead a small concentration of doped electrons leads to the formation of spin-polarons, inducing ferromagnetic order at experimentally relevant temperatures, consistent with recently observed ferromagnetic states in moiré TMD systems. We predict explicit signatures of polaron-formation in the magnetization profile. 
\end{abstract}

\date{\today}

\maketitle


Moiré heterostructures have emerged as a new platform for studying systems of strongly correlated electrons \cite{cao18,cao18b,andrei20}.
For semiconducting transition metal dichalcogenides (TMD), such as W$X_2$ ($X=\mathrm{S},\mathrm{Se}$) and Mo$X_2$ ($X=\mathrm{S},\mathrm{Te}$) \cite{wu18,wu19,tang20,kennes21}, strong spin-orbit coupling \cite{wu18,wu19}) or interlayer bonding \cite{angeli21,kaush22} lead to an effective combined spin-1/2 spin-valley degree of freedom, such that they can be modelled by single-band Hubbard models on effective moiré lattices \cite{wu18}. 
Kinetic energy scales are quenched for large moiré lattice constants, such that interactions dominate, leading to Mott-insulating behaviour at half-filling \cite{tang20,wang20} as well as at \emph{fractional} fillings \cite{xu20,regan20}, with charge configurations stabilized by effective charge-transfer potentials or due to Wigner crystallization \cite{li21,morales,padhi21} induced by longer-ranged repulsive interactions.

An important open question pertains to the magnetic states obtained by lifting of the remaining spin(-valley) degeneracy at lowest temperatures.
In the half-filled Hubbard model \cite{arovas22}, $S=1/2$ local moments with Heisenberg-type antiferromagnetic exchange interactions emerge, and various magnetic states have been predicted for moiré TMD systems \cite{pan20,motruk22}.
While an antiferromagnetic Curie-Weiss law was observed in WSe$_2$/WS$_2$ heterostructures \cite{tang20}, signatures of magnetic \emph{ordering} have remained elusive.
Interactions are strong (Ref.~\onlinecite{tang20} estimates that the onsite repulsion $U$ per kinetic energy $t$ is $U/t \approx 20$), but overall energy scales are small, such that the estimated magnetic exchange $J_\mathrm{AFM} \sim t^2/U \approx 0.05 \, \mathrm{meV}$ \cite{tang20} requires very low temperatures to observe magnetic ordering.

Magnetic ordering temperatures are expected to become even smaller in moiré-Mott insulators at dilute fillings, for example quarter-filled (one particle per moiré unit cell) moiré honeycomb lattices, such as twisted MoSe$_2$ and MoTe$_2$ or AB-stacked MoT$_2$/WSe$_2$, as well as the moiré triangular lattice of WSe$_2$/WS$_2$ at 1/3-filling \cite{li21}.
Magnetic interactions are then only induced at higher order $n>1$ in $t/U$ perturbation theory \footnote{for consistency, we assume that nearest-neighbor Coulomb interactions $V$ are on a similar order as $U$, and that longer-ranged hopping $t' \sim t^2/U$)}, rendering critical temperatures $T \sim t(t/U)^n$ experimentally inaccessible.
\begin{figure}
	\centering
	\includegraphics[width=\columnwidth]{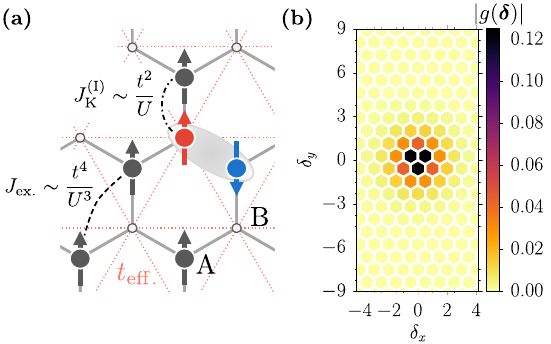}
	\caption{\label{fig:introfig}(a) Illustration of the moiré-Hubbard model at quarter filling with an additional doped electron. Local moments reside on the $\Au$ sublattice sites (blue), whereas the doped electron (red) has an effective hopping $t_\mathrm{eff.}$ between the $\Bu$ sublattice sites (red dashed lines). The Kondo exchange $\Jki$ (dash-dotted) is parametrically larger than exchange interactions $J_\mathrm{ex.}$ (dashed) between local moments, and leads to the formation of a spin polaron, a bound state between doped $\uparrow$-electron (red) and a spin-flip (blue). (b) Magnitude of the bound-state wavefunction $g(\bvec{r} = \bvec{\delta})$ in Eq.~\eqref{eq:psi-bs} for translation vectors $\bvec{\delta}$ of the triangular Bravais lattice, found by exact diagonalization in the $S^z=(N-1)/2$ sector for $\Jki/t = 1$ on a system with $48 \times 48$ unit cells and periodic boundary conditions.}
\end{figure}

In this letter, however, we show that a small density of doped electrons couples to local moments in such a dilute moiré-Mott insulator via Kondo-type interactions at order $t/U$.
This stabilizes ferromagnetism at parametrically large temperatures $T\sim t^2/U \gg t (t/U)^n$ than intrinsic exchange, constituting a novel mechanism for magnetic ordering in moiré TMDs \cite{wu18,wu19,motruk22,pan22,morales,mai23,guerci23}.

Importantly, this ferromagnetism is induced by the formation of spin-polarons form as doped electrons and magnons on top of a background of localized moments, with a spontaneous zero-field magnetization $m^z = (1-x)/2$ per unit cell, where $x=N_\text{doped}/N$ denotes density of doped electrons per unit cell.
We predict a distinct magnetization profile as an experimentally observable signature of the mechanism proposed.
Our work is directly applicable to AB-stacked WSe$_2$/MoTe$_2$, for which very recently doping-induced ferromagnetism has been observed \cite{zhao23b}, and may be of relevance to the ferromagnetic states reported in twisted MoTe$_2$ \cite{anderson23}.

{\em Moiré-Hubbard model and effective Kondo lattice.---}%
We consider the effective Hamiltonian
\begin{multline} \label{eq:h-eff-kondo}
	\mathcal{H}_{\mathrm{eff}.} = - t_\mathrm{eff.} \sum_{\langle i_\mathrm{B},j_\mathrm{B} \rangle} \left(c_{i,\sigma}^\dagger c_{j,\sigma} + \hc\right) \\+ \Jki \sum_{\langle i_\mathrm{B}, j_\mathrm{A} \rangle}\frac{1}{2} c_{i,\sigma}^\dagger \vec{\tau}_{\sigma,\sigma'} c_{i,\sigma'} \cdot \vec{S}_j  \\ +\Jkii \sum_{(i_\mathrm{B},j_\mathrm{A},l_\mathrm{B})} \left( \frac{1}{2} c_{i,\sigma}^\dagger \vec{\tau}_{\sigma,\sigma'} c_{l,\sigma'} \cdot \vec{S}_{j} + \hc \right),
\end{multline}
which describes the interaction between localized moments residing on the A-sublattice of an (effective) honeycomb lattice with itinerant electrons which disperse on the B-sublattice (kinetic energy $t_\mathrm{eff.}$) via local and itinerant Kondo interactions $\Jki$ and $\Jkii$, see also Fig.~\ref{fig:introfig}(a).
Here, $(i_\mathrm{B},j_\mathrm{A},l_\mathrm{B})$ denotes the summation over next-nearest-neighbor pairs with an intermediate A sublattice site.

We consider \eqref{eq:h-eff-kondo} as an effective model for weakly doped insulating states on the moiré honeycomb lattice at quarter-filling $\bar{n} = N_\mathrm{el.}/(2N) = 1/2$, motivated by two classes of systems:

(i)~Charge-transfer insulators in AB-stacked moiré heterobilayers such as MoTe$_2$/WSe$_2$, for which Kondo-lattice phenomenology has been observed \cite{zhao23}. Wannier centers in the two layers correspond to the A/B sublattice sites of an effective honeycomb lattice, with interlayer tunneling yielding nearest-neighbor hopping $t$. Band misalignment leads to a finite charge-transfer energy $\Delta_\mathrm{CT}$ between the A/B sublattice sites. At $\bar{n}=1/2$ and $t \ll \Delta_\mathrm{CT},U$, electrons are localized on the A sublattice sites, and doped electrons occupy the B sublattice if $U > \Delta_\mathrm{CT}$. For MoTe$_2$/WSe$_2$, DFT results estimate $U=100\, \mathrm{meV}, \Delta_\mathrm{CT} = 65 \, \mathrm{meV}$ and $(t_\mathrm{AA},t_\mathrm{BB},t_\mathrm{AB})=(4.5,9,2) \, \mathrm{meV}$ \cite{trithep22}.
The intralayer hopping $t_\mathrm{BB}$ in the WSe$_2$ layer contributes to the hopping $t_\mathrm{eff.}$ \footnote{We further note that a finite $t_\mathrm{AA} \neq 0$ leads to effective exchange interactions $J_\mathrm{AA}\sim t_\mathrm{AA}^2/U$ at quadratic order, but we expect that associated magnetic ordering will be suppressed (occurs at lower temperatures than $J_\mathrm{AA}$) due to the frustrated nature of exchange interactions on the effective triangular lattice spanned by the effective A local moments}\footnote{In our model $\mathcal{H}_\mathrm{eff.}$, derived in perturbation theory from $\mathcal{H}$ in Eq.~\eqref{eq:h-hubbard}, spin-dependent complex hopping amplitudes do not occur, but may be important for more materials-specific models \cite{trithep22,guerci23}. However, the experimentally observed doping-induced ferromagnetism of MoTe$_2$/WSe$_2$ is far from the Quantum Anomalous Hall regime \cite{zhao23b}, suggesting that spin-orbit coupling plays a negligible role in the regime of interest.}.

(ii)~Wigner-Mott insulators in twisted moiré-homobilayers such as MoSe$_2$ \cite{kaush22} as well as MoTe$_2$ \cite{anderson23} and WSe$_2$ \cite{devakul21}. Longer-ranged interactions (such as nearest neighbor repulsion $V$) can stabilize Mott-insulating generalized Wigner-crystal states \cite{regan20,kaush22} for $t \ll U,V$, where at $\bar{n}=1/2$ particles occupy one of the two sublattices of the moiré-honeycomb lattice if $U> 3V$. 
While no consensus exists for an effective model for twisted MoTe$_2$ at experimentally relevant twist angles \cite{reddy23,wang23}, DFT predictions for twisted WSe$_2$ at $\theta = 1.43^\circ$ yield $t\approx 1.2 \,\mathrm{meV}$ and $U\sim 70 \,\mathrm{meV}$ \cite{devakul21} (with $t' \lesssim 0.5 \, \mathrm{meV}$), and estimate $t\approx 0.16$-$3\,\mathrm{meV}$ with $U \sim 10 t$ for twisted MoSe$_2$ \cite{kaush22}.

The key features for systems of type (i) or (ii) can be modelled in terms the Hamiltonian
\begin{align} \label{eq:h-hubbard}
	\mathcal{H} = &-t \sum_{\substack{\langle ij \rangle\\ \sigma=\uparrow,\downarrow}} \left( c_{i,\sigma}^\dagger c_{j,\sigma} + \hc \right) + U \sum_{i} n_{i,\uparrow} n_{i,\downarrow} \nonumber\\ &+ V \sum_{\langle ij \rangle} n_i n_j - \frac{\Delta_\mathrm{CT}}{2} \sum_{i \in \text{unit cells}} (n_{i,A} - n_{i,B}).
\end{align}
We show how $\mathcal{H}_\mathrm{eff.}$ emerges from $\mathcal{H}$ at second order perturbation theory in $t\ll U, \mathrm{max}(V,\Delta_\mathrm{CT})$, and give microscopic expressions for $t_\mathrm{eff.}$, $\Jki$ and $\Jkii$ in the Supplemental Material \cite{suppl}.
Formally, $\mathcal{H}_\mathrm{eff.}$ is the exact low-energy Hamiltonian (to second order in $t \ll U,V,\Delta_\mathrm{CT}$) \emph{only if} projected to the subspace with \emph{one} doped electron on the entire B sublattice and one electron on \emph{each} A sublattice site. We will later generalize to finite doping \emph{densities}.

Under the assumptions made above, exchange interactions $J_\mathrm{ex.}$ between localized electrons occur at fourth order in $t/U \ll 1$, implying that $J_\mathrm{ex.}$ is parametrically smaller than $t_\mathrm{eff},\Jki$ and $\Jkii$.
Hence, there exists a temperature regime $J_\mathrm{ex.} \ll T \ll t_\mathrm{eff.},\Jki,\Jkii$ where $\mathcal{H}_\mathrm{eff.}$ is expected to govern. Then, the Kondo-type interactions will be crucial for lifting the spin degeneracy of the local moments and selecting magnetic order, rather than (weak) superexchange.
\begin{figure}
	\centering
	\includegraphics[width=.8\columnwidth]{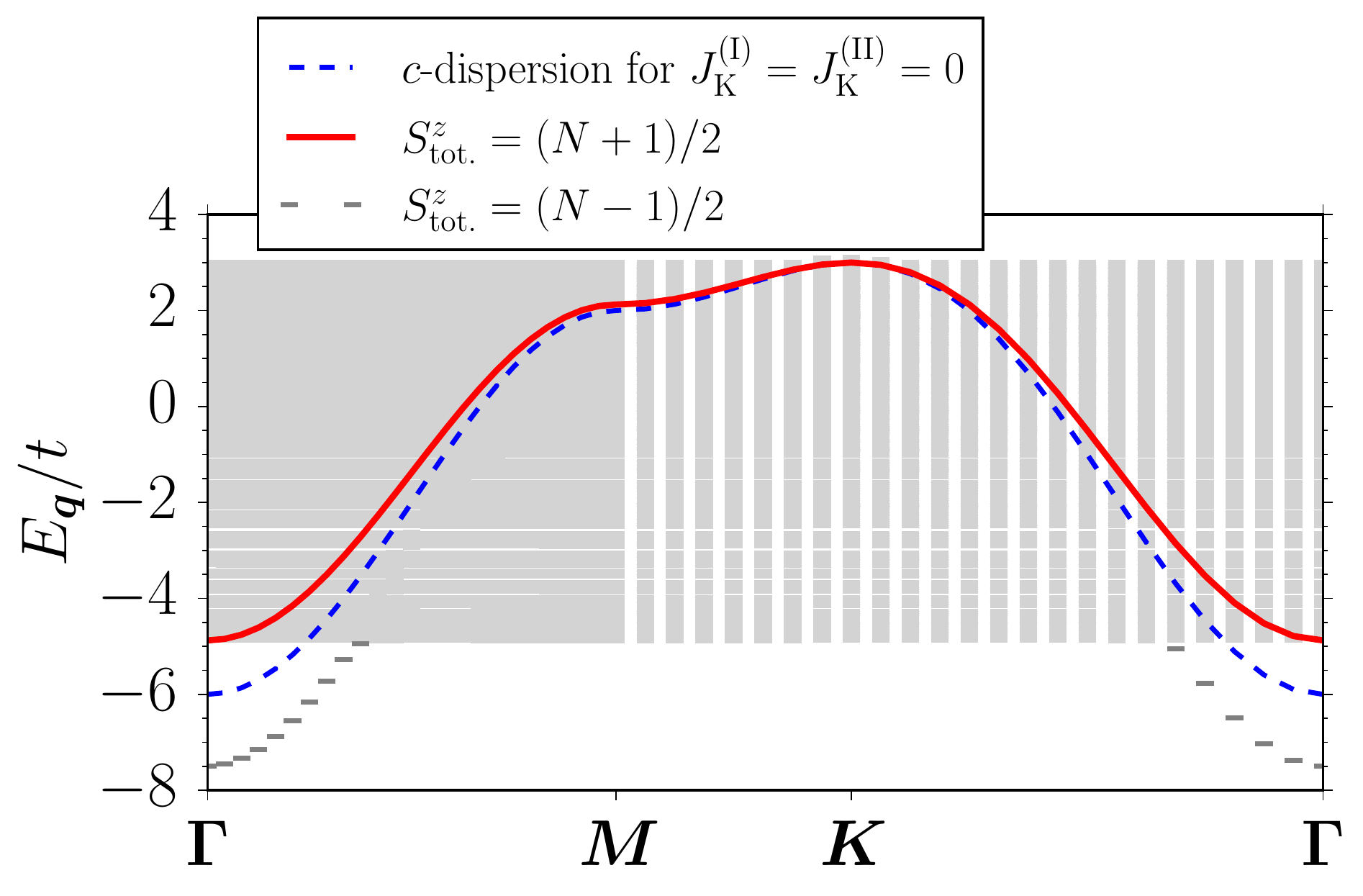}
	\caption{\label{fig:spect-pol-onedown}Spectrum of $\mathcal{H}_\mathrm{eff.}$ in Eq.~\eqref{eq:h-eff-kondo} in sectors $S^z_\mathrm{tot.} = (N+1)/2$ and $S^z_\mathrm{tot.} = (N-1)/2$ as a function of total momentum $\bvec{q}$ and $\Jki = \Jkii = 0.5t$, obtained from exact diagonalization on a finite-size cluster of $48 \times 48$ unit cells with periodic boundary conditions.}
\end{figure}

{\em Spin-polaron formation for single doped electron.---}%
To find the magnetic state selected by $\mathcal{H}_\mathrm{eff.}$, we note that the system possesses global spin rotation symmetry. The many-body Hilbert space decomposes into different sectors labelled by $S^z_\mathrm{tot.} = \sum_{i\in A} S^z_i + \sum_{i \in \mathrm{B}} c^\dagger_{i,\sigma} \tau^z_{\sigma,\sigma'} c_{i,\sigma'} /2$.
For a single doped electron, we can diagonalize $\mathcal{H}_\mathrm{eff.}$ in different sectors of $S^z_\mathrm{tot}$, and compare the respective lowest energy in each sector:
In the polarized sector with $S^z_\mathrm{tot.}=(N+1)/2$, the spectrum is given by doped $c_\uparrow$-electron dispersing in the background of polarized local moments $\ket{\Uparrow} = \prod_{i\in \mathrm{A}} \ket{\uparrow}_i$,
\begin{equation} \label{eq:disp-polarized}
	\varepsilon_{\uparrow,\Uparrow}(\bvec{k}) = \left(-2 t_\mathrm{eff.} + \frac{\Jkii}{2} \right) \Re\left[f_\mathrm{\Bu \Bu}(\bvec{k})\right] + \frac{3 \Jki}{4},
\end{equation}
where $f_\mathrm{BB}(\bvec{k}) = \eu^{\iu \bvec{k} \cdot \bvec{n}_1} + \eu^{\iu \bvec{k} \cdot \bvec{n}_2} + \eu^{\iu \bvec{k} \cdot (\bvec{n}_1-\bvec{n}_2)}$, with $\bvec{n}_1 = (1,0)^\top$ and $\bvec{n}_2 = (1,\sqrt{3})^\top/2$, and $\bvec{k}$ is a momentum in the 1. Brillouin zone (BZ) of the honeycomb lattice.

The next sector has $S^z=(N-1)/2$.
It is convenient to represent the local moments in terms of hardcore bosons (magnons) $b_i$ with $S^+_i = b_i$, $S^-_i = b_i^\dagger$ and $S^z_i = 1/2-b^\dagger_i b_i$ with respect $\ket{\Uparrow}$.
A complete basis for this is subspace is given by $\ket{\bvec{q}} = c^\dagger_{\bvec{q},\downarrow} \ket{\Uparrow}$ and $\ket{\bvec{q};\bvec{k}} = c_{\bvec{k},\uparrow}^\dagger b_{\bvec{q}-\bvec{k}}^\dagger \ket{\Uparrow}$, where $\bvec{q}\in$ 1.~BZ is the total (conserved) momentum and $\bvec{k} \in$ 1.~BZ is the relative momentum of the magnon--$\downarrow$-electron pair.
The Hamiltonian in this sector is diagonalized \cite{suppl}, and we show the spectrum as a function of $\bvec{q}$ in Fig.~\ref{fig:spect-pol-onedown}.

Importantly, there exists a dispersing band below the two-particle continuum, with a minimum at the $\Gamma$-point.
This state is understood as the doped $\downarrow$-electron hybridizing with a bound pair of a doped $\uparrow$-electron and a magnon, henceforth referred to as a \emph{spin-polaron}.
Explicitly, one may write the bound-state wavefunction (with zero external momentum $\bvec{q}=0$) as
\begin{equation} \label{eq:psi-bs}
	\ket{\psi} = \frac{1}{\mathcal{N}} \sum_{i\in \Bu} \left( c_{i,\downarrow}^\dagger + \sum_{j\in \Au} g(\bvec{r}_i - \bvec{r}_j) c_{i,\uparrow}^\dagger b_j^\dagger \right) \ket{\Uparrow},
\end{equation}
where $\mathcal{N}$ is some normalization constant, and the function $g(\bvec{r})$ can be extracted numerically, see Fig.~\ref{fig:introfig}(b), or determined perturbatively by analytical means (see below).
The bound state exists for any $\Jki/t_\mathrm{eff.},\Jkii/t_\mathrm{eff.} > 0$, and also if $\Jki/t_\mathrm{eff.} > 0,\Jkii = 0$ (and vice versa).
For simplicity, in the remainder of this work, we take $\Jkii = 0$ and focus on nonzero $\Jki /t > 0$.
Exact diagonalization in the sector of $S^z_\mathrm{tot} = (N-3)/2$ \cite{suppl} yields a spectrum featuring a symmetry-mandated descendant of $\ket{\psi}$, obtained by acting with the lowering operator $S^-_\mathrm{tot.} \ket{\psi}$, but no additional bound states emerge.

Above results suggest that the ground state of $\mathcal{H}_\mathrm{eff.}$ lies in the sector of $S^z_\mathrm{tot.} = (N-1)/2$, and features a spin-polaron (with quantum number $S^z = -1/2$) as a bound state between doped electron and a magnon, with remaining moments polarized, as has also been obtained rigorously for the one-electron Kondo chain \cite{sigrist91}.
The bound state wavefunction may be obtained analytically by using $\ket{\psi}$ in Eq.~\eqref{eq:psi-bs} as an ansatz wavefunction for the stationary Schroedinger equation, $\mathcal{H}_\mathrm{eff.} \ket{\psi} = E \ket{\psi}$.
Projecting onto the basis states for the sector $S^z_\mathrm{tot.} = (N-1)/2$ yields the coupled equations
\begin{subequations}
	\begin{align}
		\mathcal{E} &= - 6 - \frac{3 \ji}{4} + \frac{\ji}{2} \sum_{\bvec{\delta}_{\Au \Bu}} g(-\bvec{\delta_{\Au \Bu}}) \label{eq:bs-self-con-1} \\
		\tilde{g}(\bvec{q}) &\left[\mathcal{E} + 2 \Re[f_{\Bu \Bu}(\bvec{q})] - \frac{3 \ji}{4} \right] \nonumber\\ &\quad \quad= \frac{\ji}{2} \sum_{\bvec{\delta}_{\Au \Bu}} \eu^{\iu \bvec{q} \cdot \bvec{\delta}_{\Au \Bu}} \left(1 - g(-\bvec{\delta}_{\Au \Bu}) \right), \label{eq:bs-self-con-2}
	\end{align} 
\end{subequations}
with dimensionless $\mathcal{E} = E/t_\mathrm{eff.}$ and $\ji = \Jki/t_\mathrm{eff.}$.
Here, $\bvec{\delta}_{\Au \Bu} = \{0,-\bvec{n}_2,-\bvec{n}_2+\bvec{n}_1 \}$ are the translation vectors from a B sublattice site to its three neighboring A sublattice sites, and $\tilde{g}(\bvec{q}) = \sum_j \eu^{\iu \bvec{q} \cdot \bvec{r}} g(\bvec{r})$ denotes the Fourier transform of $g(\bvec{r})$.
The bound state is found to have energy $\mathcal{E} = -6  + 3\ji /4 -\delta\mathcal{E}$, where $\delta\mathcal{E}$ is the binding energy.
For $\ji \ll 1$, the self-consistency equations can be solved analytically \cite{suppl}, yielding $g(-\bvec{\delta}_{\Au \Bu}) = \mathcal{J}/2 \left(-C_0 + \sqrt{3}/(4 \pi) \ln[3 \ji/2]\right) + \mathcal{O}(\ji^2)$, where $C_0 \approx 0.256$, and the binding energy $\delta \mathcal{E} = 3 \ji /2 +\mathcal{O}(\ji^2)$.

Complementary to this perturbative analysis at $\Jki \ll t_\mathrm{eff.}$, we support our conclusions by performing a strong-coupling analysis at $\Jki \gg t_\mathrm{eff.}$, where we first focus on the Kondo interaction between the doped electron and its three neighboring A sublattice local moments, $\mathcal{H}_{\Jki} = \Jki \sum_{i=1,2,3} \vec{S}_i \cdot \frac{1}{2} c^\dagger_\sigma \vec{\tau}_{\sigma \sigma'} c_\sigma$.
The ground-state energy is minimal if the three moments form a spin-3/2 representation and the cluster of doped $c_\sigma$-electron and its three neighboring spins has total spin $S=1$, with the corresponding $S^z=1$ state constructed from $\ket{\uparrow\uparrow\uparrow}\ket{c_\downarrow}$ and $\ket{\downarrow\uparrow\uparrow + \mathrm{cycl}.}\ket{c_\uparrow}$.
The degeneracy of the remaining moments is lifted in first-order perturbation theory in the hopping of the doped electron, $\mathcal{H}_t = - t_\mathrm{eff} \sum_{\langle i_\Bu j_\Bu \rangle} (c_{i,\sigma}^\dagger c_{j,\sigma} + \hc)$: The strongly coupled clusters in the initial and final state of a hopping process between two B sublattice sites share a common spin, thus constraining matrix elements of the hopping operator projected into the degenerate ground-state manifold \cite{suppl}.
Decomposing this operator into different sectors of $S$, we find that the kinetic energy is maximal if the two local moments outside the strongly coupled cluster (involving the $c_\sigma$-electron) in the initial/final states are in a spin-1 state.
First-order perturbation theory hence selects a state with a dispersing excitation with quantum number $S = 1$, and ferromagnetic alignment of the remaining local moments.
Such a state can be achieved for $S^z_\mathrm{tot.} = (N-1)/2$, consistent with the spin-polaron in a ferromagnetic background found earlier.

{\em Ferromagnetism at finite doping: Mean-field theory and RKKY interactions.---}%
Having studied the magnetic state induced by a \emph{single} doped electron, we now consider a finite \emph{density} of doped electrons.
For sufficiently dilute doping, we neglect interactions between doped electrons, and $\mathcal{H}_\mathrm{eff.}$ [\eqref{eq:h-eff-kondo}] is applicable.
To find magnetically ordered states induced by the coupling of local moments to the dispersing doped electrons, we make a mean-field approximation for the local moments.
This approximation is controlled in the limit $\Jki \ll t_\mathrm{eff.}$, where Kondo screening effects are expected to be weak.

Assuming that the moments order in a coplanar spin spiral state with some wavevector $\bvec{Q}$, we can write $S^\pm_i \to \langle S^\pm \rangle = S \eu^{\iu \bvec{Q} \cdot \bvec{r}_i}$ and $S^z_i \to \langle S^z_i \rangle \equiv 0$.
We treat  $\bvec{Q}$ as a variational parameter for minimizing the energy of the mean-field Hamiltonian,
\begin{multline}
	\mathcal{H}^\mathrm{mf}_{\bvec{Q}} = \sum_{\bvec{k},\sigma} \xi(\bvec{k}) c_{\bvec{k}\sigma}^\dagger c_{\bvec{k}\sigma} \\
	+  \frac{\Jki S}{2} \sum_{\bvec{k}} \sum_{\bvec{\delta}_{\Au \Bu }} \left(\eu^{-\iu \bvec{Q} \cdot \bvec{\delta}_{\Au \Bu }} c_{\bvec{k}\uparrow}^\dagger c_{\bvec{k}+\bvec{Q}\downarrow} + \hc \right),
\end{multline}
where $\xi(\bvec{k}) = -2 t_\mathrm{eff.} \Re[f_{\Bu \Bu}(\bvec{k})]$.
For all $\Jki /t_\mathrm{eff.} >0$ and for various $c$-electron fillings, the energy $E_{\bvec{Q}} = \braket{ \mathcal{H}^\mathrm{mf}_{\bvec{Q}}}$ is minimal for $\bvec{Q} = 0$, corresponding to ferromagnetic order for the local moments \cite{suppl}.

This result may also be understood in terms of an RKKY indirect exchange interaction mediated by the doped $c$-electrons,
\begin{equation} \label{eq:hrkky}
\mathcal{H}_\mathrm{RKKY} = \sum_{i,j \in \mathrm{u.c.}} \sum_{\bvec{\delta}_{\Au\Bu},\bvec{\delta}'_{\Au\Bu}} J_\mathrm{RKKY}(\bvec{r}_i - \bvec{r}_j) \vec{S}_{i+\delta_{\Au \Bu}} \cdot \vec{S}_{j+\delta_{\Au \Bu}'}
\end{equation}
Here, $i,j$ index unit cells, or equivalently B sublattice sites on which $c$-electrons reside (see Ref.~\cite{suppl} for a rewriting in terms of interactions between $n$-th nearest neighbors on the $\Au$ sublattice of local moments), and $J_\mathrm{RKKY}(\bvec{r}) = (\Jki)^2 \chi_\mathrm{s}(\bvec{r})/4$ is proportional to the spin-susceptibility $\chi_\mathrm{s}$ of the dispersing $c$-electrons.
At sufficiently low doping densities, the Fermi level lies near the bottom of the band, which can therefore be approximated by a quadratic dispersion, $\xi(\bvec{k})/t_\mathrm{eff.} \approx -6 + 3 |\bvec{k}|^2 /2$.
A standard calculation for free electrons in $d=2$ spatial dimensions \cite{kittel69,fischer75} then yields the spin susceptibility $\chi_\mathrm{s}(\bvec{r}) \sim m^\ast k_\mathrm{F} \left(J_0(k_\mathrm{F} | \bvec{r}|) Y_0(k_\mathrm{F} | \bvec{r}|) + J_1(k_\mathrm{F} | \bvec{r}|) Y_1(k_\mathrm{F} | \bvec{r}|) \right)$, where $k_\mathrm{F} = |\bvec{k}_\mathrm{F}|$ is the Fermi wavevector, and $J_n,Y_m$ are the $n$-th ($m$-th) Bessel functions of first (second) kind. 
For low filling factors, the asymptotic forms of Bessel functions as $k_\mathrm{F} | \bvec{r}| \to 0$ imply that $\chi_s(\bvec{r}) \sim m^\ast k_\mathrm{F} \ln(k_\mathrm{F} | \bvec{r}|)$.
Hence, the dominant behaviour of $J_\mathrm{RKKY}(\bvec{r})$ at short distances is \emph{ferromagnetic}, consistent the result of our mean-field calculation.

{\em Field-induced magnetic transitions.---}%
We now argue that formation of spin-polarons on top of a ferromagnetic background of local moments leads to a distinct evolution of the system's magnetization as a function of an applied magnetic field $h$, which couples identically to local moments and doped electrons with $\mathcal{H}_h = - h\left(\sum_{i \in \Au} S^z_i + \sum_{j \in \Bu} (n_{i,\uparrow} - n_{i,\downarrow})/2\right)$:
At high fields, $h \gg t_\mathrm{eff.},\Jki$, both local moments and doped electrons are expected to be polarized, such that the magnetization \emph{per unit cell/local moment}, $m^z = (1/N) \sum_{i \in \Au,\Bu} \langle S^z_i \rangle$, is given by $m^z = (1+x)/2$, where we use $x=N_\mathrm{doped}/N$ to denote the doping density per unit cell.

\begin{figure}
	\centering
	\includegraphics[width=.9\columnwidth]{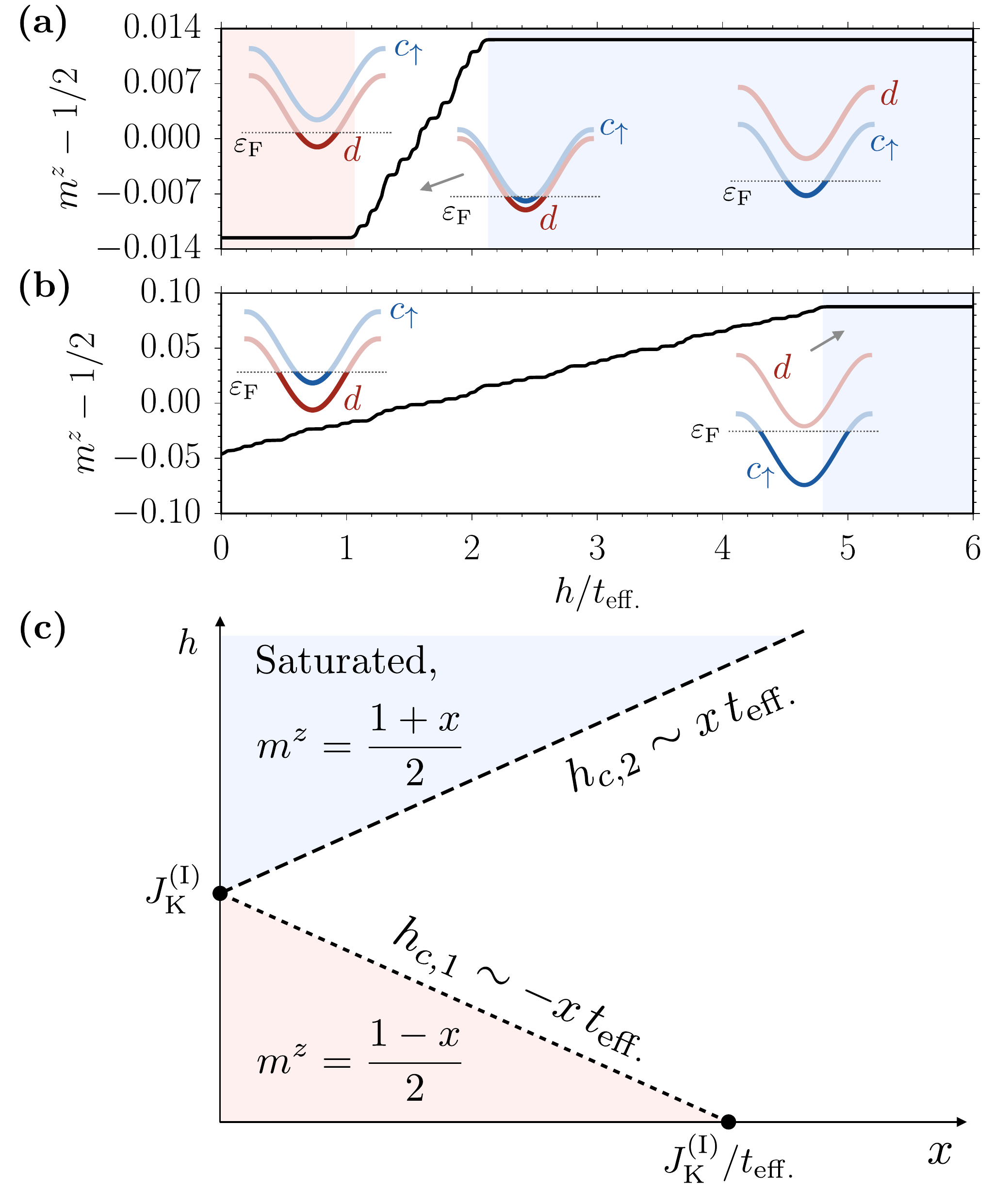}
	\caption{\label{fig:mag-h}Evolution of the magnetization $m^z$ per unit cell as a function of magnetic field $h/t_\mathrm{eff.}$ for doping densities (a) $x \approx 0.026$ and (b) $x\approx 0.17$ ($x=N_\mathrm{doped}/N$), obtained by solving the two-band model on a finite-size cluster of $48\times 48$ unit cells. The insets depict schematically the relative position and filling of $c_\uparrow$, $d$-bands. (c) Schematic phase diagram as a function of doping density $x$ and magnetic field $h$. For sufficiently small $x$, there exists a weak-field regime with a magnetization plateau at $m^z=(1-x)/2$. Note that the limit $x\to 0$ is singular.}
\end{figure}

In contrast, at $h=0$ and sufficiently dilute doping densities $x$, the ground state features  polarons hybridizing with $c_\downarrow$-electrons, on top of ferromagnetically aligned local moments.
This implies that in this limit, the magnetization will be given by
\begin{equation}
	m^z = \frac{1-x}{2}.
\end{equation}
The evolution from the $h=0$ ground state to the high-field, fully polarized, regime can be modelled in an effective two-band model, where one band is given by doped $c_\uparrow$ electrons, and the second band by spin polarons.
We construct an effective tight-binding model in real space \cite{suppl}, based on the local wavefunctions $\ket{c_{i,\uparrow}} = c_{i\uparrow}^\dagger \ket{\Uparrow}$ (where $i\in \Bu$ sublattice) and
\begin{equation}
	\ket{d_i} = \frac{\sqrt{N}}{\mathcal{N}} \left(c_{i,\downarrow}^\dagger + \sum_{j\in \Au} g(\bvec{r}_i - \bvec{r}_j) c_{i,\uparrow}^\dagger b^\dagger_{j,\downarrow}\right) \ket{\Uparrow},
\end{equation}
where $\mathcal{N}$ is the normalization constant appearing in Eq.~\eqref{eq:psi-bs}, and the function $g(\bvec{r})$ has been determined above.
The two quasiparticle bands $\varepsilon_{c_\uparrow}(\bvec{k})$ and $\varepsilon_d(\bvec{k})$ are obtained by Fourier-transforming $\braket{c_{i,\uparrow}|\mathcal{H}_\mathrm{eff.} + \mathcal{H}_h|c_{j,\uparrow}}$ and $\braket{d_{i}|\mathcal{H}_\mathrm{eff.} + \mathcal{H}_h|d_j}$.
We neglect interactions between $c_\uparrow$ and $d$-particles, and are further assume that the $d$-particles have fermionic statistics.
This is justified if the polarons are far apart so that the exponentially decaying bound-state wavefunctions $g(\bvec{r})$ have vanishing overlap, which will occur at low doping densities $x \ll 1$.

We numerically extract the function $g(\bvec{r})$ from exact diagonalization in the sector $S^z_\mathrm{tot.} = (N-1)/2$ and then evaluate the magnetization $m^z = (1/2) + 1/N \sum_{\bvec{k}} \left(\braket{c_{\bvec{k}\uparrow}^\dagger c_{\bvec{k}\uparrow}} - \braket{d_{\bvec{k}}^\dagger d_{\bvec{k}}} \right)$ on appropriately discretized momentum-space grids.
The magnetization $m^z$ as a function of $h/t_\mathrm{eff.}$ is shown for two representative doping densities $x$ in Figs.~\ref{fig:mag-h}(a)~and~(b)s \footnote{{\color{blue} We have worked small but finite temperatures $T=0.002 t_\mathrm{eff.}$ smooth out finite-size effects}}.
For small $x$, the zero-field state only features doped $c_\downarrow$-electrons/spin polarons and the Fermi level lies below the $c_\uparrow$-band, such that the magnetization is $m^z = (1-x)/2$ up to some critical field $h_{c,1}$, as shown in Fig.~\ref{fig:mag-h}(a).
Above this field, the $c_\uparrow$ band starts to become filled, and the magnetization continously evolves from the low-field to the high-field state with $m^z = (1+x)/2$, achieved at the saturation field $h_{c,2}$.
For larger doping levels [see Fig.~\ref{fig:mag-h}(b)], the zero-field splitting between $c_\uparrow$- and $d$-bands is small enough that the Fermi level lies in both bands. Then, $m^z$ immediately evolves linearly as a function of applied field to its saturation value $m^z=(1+x)/2$.
These considerations give way to the qualitative $x$-$h$ phase diagram shown in Fig.~\ref{fig:mag-h}(c).
Our results are expected to hold up to temperatures $T \sim \Jki,\Jkii$, where the spin-polarons become unstable due to thermal fluctuations.

{\em Discussion.---}%
Our key finding, that doped electrons are bound into spin-polarons on top of a ferromagnetic background of local moments, is reminiscent of Nagaoka's theorem for a hole in the half-filled Hubbard model \cite{nagaoka66, arovas22}. On the half-filled triangular lattice, doping electrons (holes) is expected to lead to (anti)-ferromagnetic correlations \cite{haerter05,zhang18,morera22,davy22}, consistent with recent experiments \cite{dehollain20,xu22,ciorciaro23}.
The geometry and interactions of the \emph{quarter}-filled system considered in our work differs significantly from these studies, resulting e.g. in different quantum numbers of the spin-polaron.
In a related work, considering Wigner-crystalized two-dimensional electron gases, Ref.~\onlinecite{kim22} has pointed out that interstitial (doped) sites contribute to ferromagnetic order \cite{hoss20}. 
Note that in moiré TMD with sufficiently strong interactions, non-local magnetic interactions can become important \cite{morales}. In particular, ferromagnetic direct exchange may aid in stabilizing the ground state discussed here.

The predicted ferromagnetic state and the magnetization curve for various fillings can be experimentally detected using optical methods \cite{cirac20}.
Indeed, very recently, ferromagnetism in quarter-filled AB-stacked MoTe$_2$/WSe$_2$ at doping densities $x=0.01$-$0.05$ has been reported \cite{zhao23b}. Magnetism is absent at $x\approx 0$, and for larger $x>0.05$ Kondo screening is dominant, strongly suggesting that magnetic ordering is intrinsically related to a finite but small doping density, consistent with our proposed mechanism.
Ferromagnetism has also been observed for twisted bilayer MoTe$_2$ at quarter-filling \cite{anderson23}.

\begin{acknowledgments}{\em Acknowledgments.}---We would like to thank Z.-X. Luo and Z.~Song for collaboration on related projects, as well as M.~Davydova, K.~F.~Mak and J.~W.~Venderbos for discussions.
UFPS is supported by the Deutsche Forschungsgemeinschaft (DFG, German Research Foundation) through a Walter Benjamin fellowship, Project ID 449890867 and the DOE office of BES, through awards number DE-SC0020305. LB is supported by the NSF CMMT program under Grants No. DMR-2116515, as well as the Simons Collaboration on Ultra-Quantum Matter, which is a grant from the Simons Foundation (651440).

\end{acknowledgments}

%

\bibliography{TMD_polaron_bib}
\bibliographystyle{apsrev4-2}

\end{document}


\title{Supplemental material:\\Spin-polarons and ferromagnetism in doped dilute Wigner-Mott insulators}

\author{Urban F. P. Seifert}
\affiliation{Kavli Institute for Theoretical Physics, University of California, Santa Barbara, CA 93106, USA}
\author{Leon Balents}
\affiliation{Kavli Institute for Theoretical Physics, University of California, Santa Barbara, CA 93106, USA}
\affiliation{Canadian Institute for Advanced Research, Toronto, Ontario, Canada}

\begin{abstract}
This Supplemental Material provides details on the derivation of $\mathcal{H}_\mathrm{eff.}$, exact diagonalization in several $S^z_\mathrm{tot.}$ sectors,  the analytical solution of the bound-state problem for a single of a single doped hole, first-order perturbation theory in the strong-coupling limit, mean-field theory and RKKY interactions, as well as the derivation of the effective two-band model. 
\end{abstract}

\date{\today}

\maketitle

\tableofcontents


\section{Derivation of the effective Hamiltonian [Eq.~(2) in main text]}

Here, we provide a derivation of the effective Hamiltonian $\mathcal{H}_\mathrm{eff.}$ [Eq.~(1) in the main text], starting from the generalized Hubbard Hamiltonian $\mathcal{H}$ [Eq.~(2), main text] using second-order perturbation theory in $t \ll U,\mathrm{max}(V,\Delta_\mathrm{CT})$. Note that it is sufficient if \emph{either} $V$ or $\Delta_\mathrm{CT}$ are non-zero for the quarter-filled reference state (with local moments residing on the A-sublattice) to be stable and thus the perturbation theory to be well-defined.

We assume that the moiré-Hubbard model [Eq.~(2) in the main text] contains only nearest-neighbor hopping amplitudes, which implies that the next-nearest neighbor hopping and Kondo interactions are real and pseudospin-rotation symmetric. For more materials-specific modelling, complex-spin-dependent second-nearest neighbor hopping amplitudes can be added to the moiré-Hubbard model, however, as long as these do not drive a band inversion (which is not the case in the regime of the experiments reported in Ref.~\onlinecite{zhao23b}), we expect our conclusions to be qualitatively robust.

\begin{figure}
	\includegraphics[width=.9\columnwidth]{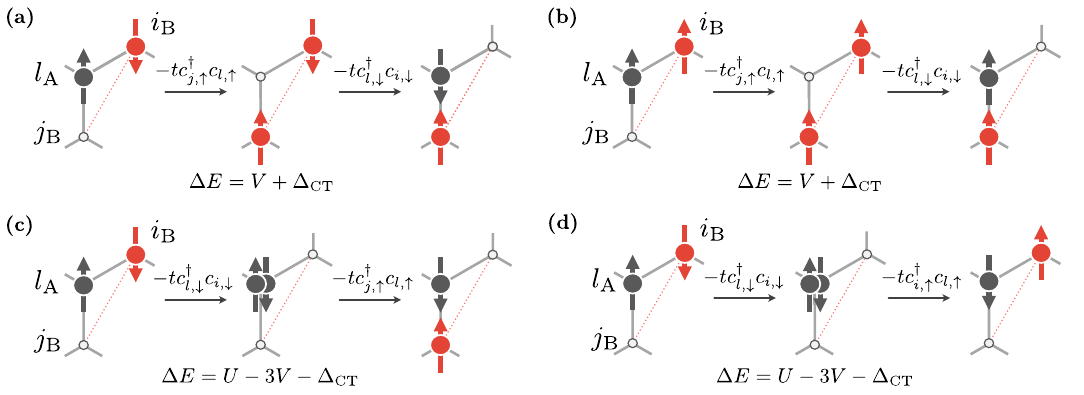}
	\caption{\label{fig:pert-theory} Illustration of processes in second-order perturbation theory that give rise to $t_\mathrm{eff.}$, $\Jki$ and $\Jkii$ at second-order perturbation theory. (a) Hopping process \emph{without} virtual doublon excitations, yielding contributions to $\mathrm{t}_\mathrm{eff},\Jkii \sim t^2/(V+\Delta_\mathrm{CT})$. (b) Spin-flip assisted-hopping without doublon-excitation, resulting in a contribution to $\Jkii \sim t^2/(V+\Delta_\mathrm{CT})$. (c) Doublon-assisted hopping, giving rise to $\Jkii \sim t^2 / (U-3V-\Delta_\mathrm{CT})$. (d) ``Local'' Kondo exchange process which leads to $\Jki \sim t^2 / (U-3V-\Delta_\mathrm{CT})$. Note that the process (not shown) where a doublon is created at site $i_\mathrm{B}$ involves a virtual excitation with energy $\Delta E = U + V+\Delta_\mathrm{CT}$, corresponding to an additional term $\Jki \sim t^2/(U+V+\Delta_\mathrm{CT})$.}
\end{figure}

\subsection{Doublon-free hopping} \label{sec:doub-free}

At second order-perturbation theory, there exist matrix elements for the doped particle with spin $\sigma$ to hop from site $i_\mathrm{B}$ to site $j_\mathrm{B}$ (next-nearest neighbors on the honeycomb lattice) \emph{without} the virtual excitation of a doublon, see Figs.~\ref{fig:pert-theory}(a) and (b).
Note that the process depends on the spin of the localized electron on site $l_\mathrm{A}$. Using second-order perturbation theory, we have
\begin{equation} \label{eq:2nd-order}
	\braket{(j_\mathrm{B},\sigma'),(l_\Au,\tau)|\mathcal{H}_\mathrm{eff}|(i_\mathrm{B},\sigma),(l_\Au,\sigma_\Au)} = - \sum_m \frac{\braket{(j_\mathrm{B},\sigma'),(l_\Au,\sigma_\Au')|\mathcal{H}_t|m}\braket{m| \mathcal{H}_t| (i_\mathrm{B},\sigma),(l_\Au,\sigma_\Au)}}{E_m - E_0},
\end{equation}
where $\ket{m}$ denotes intermediate states with both $i_\Bu$ and $j_\Bu$ occupied, but no particle residing on the $l_\Au$ site, which incur an energy cost $E_m - E_0 = V + \Delta_\mathrm{CT}$. Given the $\SUtwo$-conserving nature of nearest-neighbor hopping $\mathcal{H}_t = -t \sum_{\langle m_\Au, n_\Bu\rangle} (c_{m,\sigma}^\dagger c_{n,\sigma} + \hc)$, it becomes clear that $\sigma = \sigma'$ requires $\sigma = \sigma_\Au = \sigma_\Au'$, and $\sigma'=-\sigma'$ can occur for $\sigma' = \sigma_\Au = -\sigma_\Au'$. Hence, for each initial and final state, there exists exactly one intermediate state $m$.
The resulting contribution to $\mathcal{H}_\mathrm{eff}$ can be written in the form
\begin{align}
	\mathcal{H}_\mathrm{eff} &\sim (-1)^2\frac{t^2}{V+\Delta_\mathrm{CT}} \sum_{(j_\Bu l_\Au i_\Bu)} \left[ c_{j,\uparrow}^\dagger \left(\frac{1}{2} + S^z_l \right) c_{i,\uparrow} + c_{j,\downarrow}^\dagger \left(\frac{1}{2} - S^z_l \right) c_{i,\downarrow} + c_{j,\downarrow}^\dagger S^+_l c_{i,\uparrow} + c_{j,\uparrow}^\dagger S^-_l c_{i,\downarrow} + \hc  \right] \nonumber\\
	&= \frac{t^2}{V+\Delta_\mathrm{CT}} \sum_{(j_\Bu l_\Au i_\Bu)} \left[\frac{1}{2} c_{i,\sigma}^\dagger c_{j,\sigma} + 2 \left(\frac{1}{2} c^\dagger_{j,\sigma}\vec{\tau}_{\sigma,\sigma'} c_{i,\sigma'} \right) \cdot \vec{S}_l + \hc\right], \label{eq:doub-free}
\end{align}
where $\tau^\alpha$, $\alpha = x,y,y$ denote the three Pauli matrices. Note that the additional minus sign arises due to fermionic anticommutation relations. Specifically, we pick the ordering $\ket{(i,\uparrow),(i,\downarrow),(l,\uparrow),\dots,(j,\downarrow)} = c_{i\uparrow}^\dagger c_{i\downarrow}^\dagger c_{l\uparrow}^\dagger \dots c_{j\downarrow}^\dagger$ (but any \emph{consistent} choice will be equivalent), and then, e.g., $c_{j,\uparrow}^\dagger c_{i,\downarrow} \ket{(i,\downarrow),(l,\uparrow)} = c_{j,\uparrow}^\dagger c_{i,\downarrow} c_{i,\downarrow}^\dagger c_{l,\uparrow}^\dagger \ket{0} = - c_{l,\uparrow}^\dagger c_{j,\uparrow}^\dagger \ket{0} = - \ket{(j,\downarrow),(l,\uparrow)}$). 

\subsection{Doublon-assisted hopping}

We now consider processes where the particle at $i_\Bu$ with spin $\sigma$ hops to site $j_\Bu$ (spin $\sigma'$) with a virtual doublon excitation at site $l_\Au$, see also Fig.~\ref{fig:pert-theory}(c).
Note that such process (by Pauli-exclusion) is only allowed if the local moment at site $l_\Au$ has initially spin $\sigma_\Au=-\sigma$. We again employ Eq.~\eqref{eq:2nd-order}, and the relevant matrix elements of $\mathcal{H}_t$ enter as
\begin{align}
	\left(-t \sum_{\sigma} c_{l,\sigma}^\dagger c_{i,\sigma} \right) \ket{(i_\Bu,\uparrow),(l_\Au,\downarrow)} &= - t \left( - \ket{(l_\Au,\downarrow),(j_\Bu,\uparrow)} + \ket{(l_\Au,\uparrow),(j_\Bu,\downarrow)} \right) \\ 
	\left(-t \sum_{\sigma} c_{l,\sigma}^\dagger c_{i,\sigma}\right) \ket{(i_\Bu,\downarrow),(l_\Au,\uparrow)} &= - t \left( \ket{(l_\Au,\downarrow),(j_\Bu,\uparrow)} + \ket{(l_\Au,\downarrow),(j_\Bu,\uparrow)} \right). 
\end{align}
The intermediate state with a doublon on site $l_\Au$ and no particles on any $\Bu$-sublattice sites has an energy difference to the ground state $E_m - E_0 = U - 3V-\Delta_\mathrm{CT}$ and thus the contribution to $\mathcal{H}_\mathrm{eff}$ can be written as
\begin{equation}
	\mathcal{H}_\mathrm{eff} \sim \frac{t^2}{U-3V-\Delta_\mathrm{CT}} \sum_{(j_\Bu,l_\Au,i_\Bu)} \left[ -\frac{1}{2} c_{i,\sigma}^\dagger c_{j,\sigma} + 2 \left(\frac{1}{2} c^\dagger_{j,\sigma}\vec{\tau}_{\sigma,\sigma'} c_{i,\sigma'} \right) \cdot \vec{S}_l + \hc\  \right] \label{eq:doub-assist}
\end{equation}

\subsection{Local doublon-mediated exchange}

A local Kondo-type exchange interaction is obtained if the particle at site $i_\Bu$ hops to site $l_\Au$, thereby creating a doublon at site $l_\Au$, or vice versa, see also Fig.~\ref{fig:pert-theory}(d) for an example.
The intermediate state with a doublon on site $l_\Au$ has an energy $\Delta E = U-3V - \Delta_\mathrm{CT}$ on top of the ground state, while a doublon on site $i_\Bu$ can be seen to have energy $\Delta E = U + V+\Delta_\mathrm{CT}$ on top of the ground state.
The calculation proceeds as above, and the respective contributions to the effective Hamiltonian read
\begin{equation}
	\mathcal{H}_\mathrm{eff} \sim \left(\frac{t^2}{U-3V-\Delta_\mathrm{CT}} + \frac{t^2}{U+V+\Delta_\mathrm{CT}} \right) \sum_{\langle i_\Bu,l_\Au \rangle} \left[2 \left(\frac{1}{2} c^\dagger_{i,\sigma}\vec{\tau}_{\sigma,\sigma'} c_{i,\sigma'} \right) \cdot \vec{S}_l - \frac{1}{2} c_{i,\sigma}^\dagger c_{i,\sigma}  \right]. \label{eq:kondo}
\end{equation}
We note that for $V=\Delta_\mathrm{CT}=0$ and projecting to the subspace where all $i_\Bu$ sites are half-filled, we obtain the conventional Heisenberg exchange interaction with $J_\mathrm{H} = 4 t^2 /U$.

\subsection{Definition of effective hopping and Kondo couplings}

From Eqs.~\eqref{eq:doub-free}, \eqref{eq:doub-assist} and \eqref{eq:kondo}, we can now read off the effective parameters in Eq.~(2) of the main text,
\begin{align}
	t_\mathrm{eff.} = &\frac{1}{2} \left(\frac{t^2}{U-3V-\Delta_\mathrm{CT}} - \frac{t^2}{V+\Delta_\mathrm{CT}} \right) \quad \Jki = \frac{2 t^2}{U-3V-\Delta_\mathrm{CT}} + \frac{2t^2}{U+V+\Delta_\mathrm{CT}} \nonumber\\
	&\quad \text{and} \quad \Jkii = \frac{2t^2}{V+\Delta_\mathrm{CT}} + \frac{2 t^2}{U-3V-\Delta_\mathrm{CT}},
\end{align}
and we are free to redefine the chemical potential as we work at fixed fillings.

\section{Diagonalization of $\mathcal{H}_\mathrm{eff.}$ in different $S^z_\mathrm{tot.}$ sectors}

Using the hard-core boson representation for the spin operators, the Hamiltonian in real space reads
\begin{multline}
	\mathcal{H}_\mathrm{eff.} = - t_\mathrm{eff.} \sum_{\langle i_\Au j_\Au \rangle} \left( c_{i,\sigma}^\dagger c_{j,\sigma} + \hc \right) + \Jki \sum_{\langle i_\Bu,j_\Au \rangle} \left( \frac{1}{2} c_{i,\uparrow}^\dagger c_{i,\downarrow} b_j^\dagger + \frac{1}{2} c_{i,\downarrow}^\dagger c_{i,\uparrow} b_j + \frac{n_{i,\uparrow} - n_{i,\downarrow}}{4} - \frac{n_{i,\uparrow} - n_{i,\downarrow}}{2} n_j^{(b)} \right) \\
	+ \Jkii \sum_{(i_\Bu,j_\Au,l_\Bu)} \left( \frac{1}{2} c_{i,\uparrow}^\dagger c_{l,\downarrow} b_j^\dagger + \frac{1}{2} c_{i,\downarrow}^\dagger c_{l,\uparrow} b_j + \frac{c_{i,\uparrow}^\dagger c_{l,\uparrow} - c_{i,\downarrow}^\dagger c_{l,\downarrow}}{4} + \frac{c_{i,\uparrow}^\dagger c_{l,\uparrow} - c_{i,\downarrow}^\dagger c_{l,\downarrow}}{2} n_j^{b} + \hc \right).
\end{multline}

\subsection{Subspace with $S^z_\mathrm{tot.}=(N-1)/2$}

The sector of $S^z_\mathrm{tot.} = (N -1)/2$ is spanned by (we now change notation and use $i,j$ to label unit cells, such that $c_{i,\sigma} \equiv c_{(i,\Bu),\sigma}$ denote the doped electron on the $\Bu$ site of unit cell $i$)
\begin{equation} \label{eq:sz-basis-states}
	\ket{(i,\downarrow)} \equiv c_{i,\downarrow}^\dagger \ket{0} \quad \text{and} \quad \ket{(i,\uparrow);(-1)_j} \equiv c_{i,\uparrow}^\dagger b_j^\dagger \ket{0},
\end{equation} where the latter notation indicates that a spin-flip has occured on the $\Au$-sublattice site in unit cell $j$.
For $N=N_\mathrm{sites}/2$ unit cells, this Hilbert-space (in real space) is $N(N+1)$-dimensional and thus diagonalizing $\mathcal{H}_\mathrm{eff.}$ in this sector requires diagonalizing a $N(N+1) \times N(N+1)$ matrix.

We can simplify the problem by exploiting the translational symmetry of the lattice, so that we can Fourier-transform $c_i = \frac{1}{\sqrt{N}} \sum_{\bvec{k}} \eu^{\iu \bvec{k} \cdot \bvec{r}_i} c_{\bvec{k}}$, and the first part of the Hamiltonian becomes
\begin{equation}
	\mathcal{H}_t = -t_\mathrm{eff.} \sum_{\bvec{q},\sigma} 2 \Re[f_{\Bu \Bu}(\bvec{q})] c_{\bvec{q},\sigma}^\dagger c_{\bvec{q},\sigma}.
\end{equation}
We also introduce the collection of vectors (indexed by $\alpha =1,2,3$)
\begin{equation} \label{eq:delta-AB}
	\bvec{\delta}_{\Au \Bu}^{(\alpha)} = \{0,-\bvec{n}_2,-\bvec{n}_2 + \bvec{n}_1\}
\end{equation}
such that $i+\delta_{\Au \Bu}^{(\alpha)}$ are the unit cells of the three A sublattice sites neighboring the site $(i,\Bu)$,  and
\begin{equation} \label{eq:delta-BAB}
	\bvec{\delta}_{\Bu (\Au) \Bu}^{(\alpha)} = \{\bvec{n}_1 - \bvec{n}_2,0,0 \}
\end{equation}
are unit vectors that refer to the A sublattice site through which effective hopping between two $B$ sublattice sites occurs. Note that it is paramount that the same ordering in Eqs.~\eqref{eq:delta-AB} and \eqref{eq:delta-BAB} is maintained.
Then, one can write in momentum space
\begin{multline}
	\mathcal{H}_{\Jki}/\Jki = \frac{1}{2\sqrt{N}} \sum_{\bvec{q},\bvec{k}} \left( \rho^\dagger_q(\bvec{k}) c_{\bvec{q},\downarrow} f_{\Au \Bu} (\bvec{k}-\bvec{q}) + c_{\bvec{q},\downarrow}^\dagger \rho_q(\bvec{k}) f_{\Au \Bu}^\ast(\bvec{k}-\bvec{q}) \right) + \frac{3}{4} \sum_{\bvec{q}} \left(c_{\bvec{q},\uparrow}^\dagger c_{\bvec{q},\uparrow} - c_{\bvec{q},\downarrow}^\dagger c_{\bvec{q},\downarrow} \right)\\
	\qquad- \frac{1}{2N}\sum_{\bvec{p},\bvec{k},\bvec{q}} \left( \rho_{\bvec{p}}^\dagger(\bvec{k}) \rho_{\bvec{p}}(\bvec{q}) f_{\Au \Bu }(\bvec{k}-\bvec{q}) - c_{\bvec{k},\downarrow}^\dagger c_{\bvec{q},\downarrow} b^\dagger_{\bvec{p}-\bvec{k}} b_{\bvec{p}-\bvec{q}}  f_{\Au \Bu}(\bvec{k}-\bvec{q})  \right),
\end{multline}
where we have introduced the operator $\rho_{\bvec{q}}^\dagger(\bvec{k}) = c^\dagger_{\bvec{k},\uparrow} b^\dagger_{\bvec{q}-\bvec{k}}$ and the function $f_{\Au \Bu}(\bvec{k}) = \sum_\alpha \eu^{\iu \bvec{\delta}_{\Au \Bu}^{(\alpha)} \cdot \bvec{k}}$.
Similarly, the second part of the interaction reads
\begin{multline}
\mathcal{H}_{\Jkii} / \Jkii = \frac{1}{{2}\sqrt{N}} \sum_{\bvec{q},\bvec{k}} \sum_\alpha \left[ \rho_{\bvec{q}}^\dagger(\bvec{k}) c_{\bvec{q},\downarrow} \eu^{\iu \bvec{q} \cdot  \bvec{\delta}_{\Bu \Bu}^{(\alpha)} + \iu (\bvec{k}-\bvec{q}) \cdot \bvec{\delta}^{(\alpha)}_{\Bu(\Au)\Bu} } + c_{\bvec{q},\downarrow}^\dagger \rho_{\bvec{q}}(\bvec{k}) \eu^{\iu \bvec{q} \cdot \delta_{\Bu \Bu}^{(\alpha)} + \iu (\bvec{k}-\bvec{q}) \delta_{\Bu(\Au)\Bu}^{(\alpha)}} + \hc \right]  \\
	 +\frac{1}{4} \sum_{\bvec{q}} 2 \Re[f_{\Bu \Bu }(\bvec{q})] \left( c_{\bvec{q},\uparrow}^\dagger c_{\bvec{q},\uparrow} - c_{\bvec{q},\downarrow}^\dagger c_{\bvec{q},\downarrow} \right) - \frac{1}{2N} \sum_{\bvec{k},\bvec{p},\bvec{q}} \sum_\alpha \left[ \left( \rho_{\bvec{p}}^\dagger(\bvec{k}) \rho_{\bvec{p}}(\bvec{q}) + c_{\bvec{k},\downarrow}^\dagger c_{\bvec{q},\downarrow} b^\dagger_{\bvec{p}-\bvec{k}} b_{\bvec{p}-\bvec{q}} \right) \eu^{\iu \bvec{q} \cdot \bvec{\delta}_{\Bu \Bu}^{(\alpha)} + \iu \delta_{\Bu(\Au)\Bu}^{(\alpha)} \cdot (\bvec{k}-\bvec{q})} + \hc \right].
\end{multline}
The spectrum is then obtained by solving the $N$ eigenproblems (for each $\bvec{q}$) $\underline{\mathcal{H}}_{\bvec{q}} \underline{u}_{\bvec{q}} = E_{\bvec{q}} \underline{u}_{\bvec{q}}$ with the $(N+1) \times (N+1)$-dim. matrix
\begin{equation}
	\underline{\mathcal{H}}_{\bvec{q}} = \begin{pmatrix}
		\braket{\bvec{q}|\mathcal{H}|\bvec{q}} & \braket{\bvec{q}|\mathcal{H}|\bvec{q};\bvec{k}_1} & \dots & \braket{\bvec{q}|\mathcal{H}|\bvec{q};\bvec{k}_N} \\
		\braket{\bvec{q};\bvec{k}_1|\mathcal{H}|\bvec{q}} & \braket{\bvec{q};\bvec{k}_1|\mathcal{H}|\bvec{q};\bvec{k}_1} & \dots & \braket{\bvec{q};\bvec{k}_1|\mathcal{H}|\bvec{q};\bvec{k}_N} \\
		\vdots & \vdots & \vdots & \vdots \\
		\braket{\bvec{q};\bvec{k}_N|\mathcal{H}|\bvec{q}} & \braket{\bvec{q};\bvec{k}_N|\mathcal{H}|\bvec{q};\bvec{k}_1} & \dots & \braket{\bvec{q};\bvec{k}_N|\mathcal{H}|\bvec{q};\bvec{k}_N} \label{eq:bloch-mat}
	\end{pmatrix},
\end{equation}
where we denote the states $\ket{\bvec{q}} = c_{\bvec{q},\downarrow}^\dagger \ket{\Uparrow}$ and $\ket{\bvec{q};\bvec{k}}= \rho^\dagger_{\bvec{q}}(\bvec{k}) \ket{\Uparrow}$.

\subsection{Subspace with $S^z_\mathrm{tot.} = (N-3)/2$}

The sector $S^z_\mathrm{tot.} = (N-3)/2$ is spanned by states with a doped spin-down electron and a single spin-flip, or a doped spin-up electron and two spin flips. 
Because the spin flips are represented by hard-core bosons, states with double occupancy must be projected out, which requires us to work in real space.
Moreover, guided by the fact that the bound state above occured at $\bvec{q}=0$, we will work at $\bvec{q}=0$ external momentum.
Then, we can define the states
\begin{equation}
	\ket{\downarrow;\delta} = \frac{1}{\sqrt{N}} \sum_{i} c^\dagger_{i,\downarrow} b_{i+\delta}^\dagger \ket{\mathrm{FM}},
\end{equation}
where $\delta$ is any lattice vector of the hexagonal Bravais lattice (again we can suppress sublattice indices by noting that $c,b$ are confined to B, A sublattices).
Moreover, we can define
\begin{equation} \label{eq:d1d2}
	\ket{\uparrow;\delta_1,\delta_2} = \frac{1}{\sqrt{N}} \sum_{i} c^\dagger_{i,\uparrow} b_{i+\delta_1}^\dagger b_{i+\delta_2}^\dagger \ket{\mathrm{FM}}.
\end{equation}
Focussing, for simplicity, one the case of $\Jki \neq 0$ and $\Jkii = 0$,
we can now write down the non-zero matrix elements of the Hamiltonian $\mathcal{H}_\mathrm{eff}$ with respect to above states, which are given by
\begin{equation}
	\braket{\downarrow;\delta+\delta_{\Bu \Bu}|\mathcal{H}_t|\downarrow;\delta} = -t \quad \text{and} \quad \braket{\downarrow;\delta-\delta_{\Bu \Bu}|\mathcal{H}_t|\downarrow;\delta} = -t,
\end{equation}
\begin{equation}
	\braket{\uparrow;\delta+\delta_{\Bu \Bu},\delta'+\delta_{\Bu \Bu}|\mathcal{H}_t|\uparrow;\delta,\delta'} = -t \quad \text{and} \quad \braket{\uparrow;\delta-\delta_{\Bu \Bu},\delta'-\delta_{\Bu \Bu}|\mathcal{H}_t|\uparrow;\delta,\delta'} = -t,
\end{equation}
where $\delta_{\Bu \Bu}$ can take $\delta_{\Bu \Bu} = \{\bvec{n}_1,\bvec{n}_2,\bvec{n}_2-\bvec{n}_1\}$.
These correspond to the hopping of the $c$-electron in a comoving reference frame.
Next, the Kondo interaction gives rise to
\begin{equation}
	\braket{\uparrow;\delta,\delta_{\Au \Bu}|\mathcal{H}_{\Jki}| \downarrow;\delta} = \frac{\Jki}{2} \quad \text{and} \quad \hc,
\end{equation}
where $\delta_{\Au \Bu} = \{0,-\bvec{n}_2,-\bvec{n}_2+\bvec{n}_1\}$.
Finally, the longitudinal matrix elements are given by
\begin{align}
	\braket{\downarrow;\delta |\mathcal{H}_{\Jki}| \downarrow;\delta} &= - \frac{3 \Jki}{4} + \frac{\Jki}{2} \sum_{\delta_{\Au \Bu}} \delta_{\delta,\delta_{\Au \Bu }} \\
	\braket{\uparrow;\delta,\delta'|\mathcal{H}_{\Jki}| \uparrow;\delta,\delta'} &= \frac{3 \Jki}{4} -\frac{\Jki}{2} \sum_{\delta_{\Au \Bu}} \left( \delta_{\delta,\delta_{\Au \Bu}} + \delta_{\delta',\delta_{\Au \Bu}} \right)
\end{align}

\begin{figure}
	\centering
	\includegraphics[width=.5\textwidth]{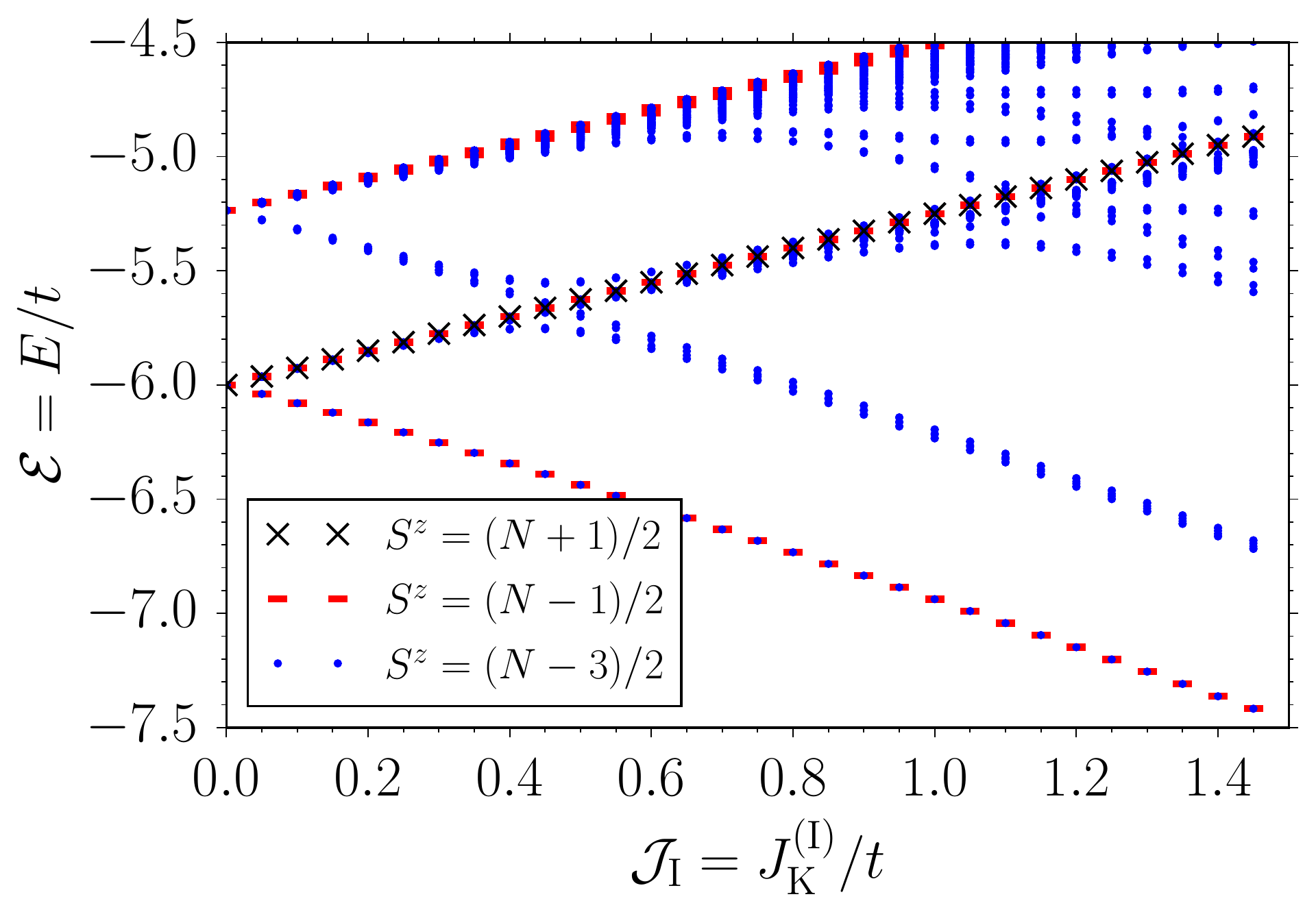}
	\caption{\label{fig:two-down}Spectra of the Hamiltonian $\mathcal{H}_\mathrm{eff.}$ at total external momentum $\bvec{q}=0$ for three distinct sectors of $S^z_\mathrm{tot.}$, obtained on a finite-size cluster of $10 \times 10$ unit cells.}
\end{figure}

We again stress that states with two bosons on a single site are to be projected out, leaving $N(N-1)/2$ distinct two-magnon states.
The Hamiltonian is then diagonalizing numerically for $\delta, \delta'$ on a lattice of $10 \times 10$ (with periodic boundary conditions), and we show the spectrum as a function of $\Jki/t_\mathrm{eff.}$ in Fig.~\ref{fig:two-down}, along with the spectra at total momentum $\bvec{q}=0$ in the $S^z_\mathrm{tot.} = \frac{N+1}{2}$ and $S^z_\mathrm{tot.} = \frac{N-1}{2}$ sectors.
As is clearly visible, the spectrum in the $S^z_\mathrm{tot.} = \frac{N-3}{2}$ contains the $\SUtwo$-symmetry-mandated copies of the spectra in the higher-weight sectors (the corresponding eigenstates are constructed from acting with $S^-_\mathrm{tot.}$ on states in the $S^z_\mathrm{tot} = (N+1)/2$ and $S^z_\mathrm{tot} = (N-1)/2$ sectors, respectively).
Importantly, there does \emph{not} exist a new bound state below the lowest-energy levels (bound state) found in the $S^z_\mathrm{tot.} = (N-1)/2$ sector.

\section{Analytical solution of self-consistency relations for bound-state wavefunction}

We obtain the self-consistency equations (5a), (5b) in the main text for the Fourier coefficients $\tilde{g}(\bvec{q})$ from the Schroedinger equation $\mathcal{H}_\mathrm{eff.} \ket{\psi} = E \ket{\psi}$ by projecting onto the states $\ket{\downarrow} = \mathcal{N}^{-1} \sum_{i \in \Bu} c_{i,\downarrow}^\downarrow$ and $\ket{(i,j);\uparrow} = \mathcal{N}^{-1} c_{i,\uparrow}^\dagger b_{j}^\dagger \ket{\Uparrow}$, yielding
\begin{equation}
	\braket{\downarrow|\mathcal{H}_\mathrm{eff.}|\psi} = N E \quad \text{and} \quad \braket{(i,j);\uparrow} = g(\bvec{r}_i -\bvec{r}_j) E.
\end{equation}
Fourier-transforming then leads to Eqs.~(5a) and (5b) in the main text, which we now solve perturbatively.
To this end, we first got to the continuum limit where $N^{-1} \sum_{\bvec{q} \in\mathrm{BZ}} \to \Omega (2\pi)^{-2} \int_\mathrm{BZ} \du^2 \bvec{q}$ with the unit-cell volume $\Omega = \sqrt{3}/2$.
From Eq.~(5b) in the main text, we get the real-space form $g(\bvec{r})$ as
\begin{equation}
	g(\bvec{r}) = \frac{\ji}{2} \sum_{\bvec{\delta}_{\Au \Bu}} (1-g(-\bvec{\delta}_{\Au \Bu})) \times \Omega \int_{\mathrm{BZ}} \frac{\du^2 \bvec{q}}{(2\pi)^2} \frac{\eu^{\iu \bvec{q} \cdot (\bvec{\delta}_{\Au \Bu} +\bvec{r})}}{\mathcal{E} + 2 \Re[f_{\Bu \Bu}(\bvec{q})] - \frac{3 \ji}{4}}.
\end{equation}
Evaluating $g(\bvec{r})$ at $\bvec{r} = -\bvec{\delta}_{\Au \Bu}^{(\alpha)}$ then leads to a coupled system of implicit equations which can be written in matrix form as
\begin{equation} \label{eq:bs-matrix-form}
	\begin{pmatrix}
		\frac{2}{\mathcal{J}_\mathrm{I}} + I(0) & I_{2,1} & I_{3,1} \\
		 I_{2,1}^\ast & \frac{2}{\mathcal{J}_\mathrm{I}} + I(0) & I_{3,2} \\
		 I_{3,1}^\ast & I_{3,2}^\ast & \frac{2}{\mathcal{J}_\mathrm{I}} + I(0)
	\end{pmatrix} \begin{pmatrix}
		g_1 \\ g_2 \\ g_3
	\end{pmatrix} = \begin{pmatrix}
		I(0) + I_{2,1} + I_{3,1} \\
		I(0) + I_{2,1}^\ast + I_{3,2} \\
		I(0) + I_{3,1}^\ast + I_{3,2}^\ast
	\end{pmatrix},
\end{equation}
where $g_\alpha \equiv g(-\bvec{\delta}_{\Au \Bu}^{(\alpha)})$ and we define the integrals $I_{\beta,\alpha} \equiv I(\bvec{\delta}_{\Au \Bu}^{(\beta)} - \bvec{\delta}_{\Au \Bu}^{(\alpha)})$ with
\begin{equation} \label{eq:i-delta}
	I(\bvec{\delta}) = \Omega \int_\mathrm{BZ} \frac{\du^2 \bvec{q}}{(2\pi)^2} \frac{\eu^{\iu \bvec{q} \cdot \bvec{\delta}}}{\mathcal{E} + 2 \Re[f_{\Bu \Bu}(\bvec{q})] - \frac{3 \ji}{4}},
\end{equation}
with the poles determining the the energy of the continuum with $\mathcal{E}(\bvec{q}) = -2 \Re[f_{\Bu \Bu}(\bvec{q})] + 3 \ji/4$ (recall that we focus on $\bvec{k}=0$ external momentum).
Noticing that $6 \geq 2 \Re[f_{\Bu \Bu}(\bvec{q})]$ seek bound states outside the continuum which have
\begin{equation} \label{eq:e-bs}
	\mathcal{E} = - 6 + \frac{3 \mathcal{J}_\mathrm{I}}{4} - \delta \mathcal{E}
\end{equation}
 with $\delta \mathcal{E}> 0$.
In this case, \eqref{eq:i-delta} becomes singular as $\delta \mathcal{E} \to 0^+$ and we can focus on its asymptotic behaviour.

To this end, we note that inserting \eqref{eq:e-bs} in the denominator in \eqref{eq:i-delta} yields
\begin{equation}
	\frac{1}{\mathcal{E} + 2\Re[f_{\Bu \Bu }(\bvec{q})] - \frac{3 \ji }{4}} = -\frac{1}{f(\bvec{q}) + \delta \mathcal{E}}
\end{equation}
where we introduce some function $f(\bvec{q})$. The series expansion 
\begin{equation} \label{eq:f_series_expansion}
	f(\bvec{q}) = 6 - 2 \Re[f_{\Bu \Bu}(\bvec{q})] = 3 |\bvec{q}|^2 /2 - 3 |\bvec{q}|^4 /32 + \dots
\end{equation}
will be useful lateron. Focussing on $\bvec{\delta}=0$ for concreteness, we can now extract the asymptotic behaviour by adding and subtracting the divergent contribution to the integral,
\begin{multline}
	-\frac{I(0)}{\Omega} = \int_\mathrm{BZ} \frac{\du^2 \bvec{q}}{(2 \pi)^2} \left[\frac{1}{f(\bvec{q}) + \delta \mathcal{E}} - \frac{1}{3 |\bvec{q}|^2/2 +\delta\mathcal{E}} + \frac{1}{3 |\bvec{q}|^2/2 +\delta\mathcal{E}} \right] = \int_\mathrm{BZ} \frac{\du^2 \bvec{q}}{(2 \pi)^2} \left[ \frac{\frac{3}{2}|\bvec{q}|^2 + \delta \mathcal{E} - (f(\bvec{q}) + \delta \mathcal{E})}{\left(\frac{3}{2}|\bvec{q}|^2 + \delta \mathcal{E}\right)(f(\bvec{q}) + \delta \mathcal{E})} + \frac{1}{\frac{3}{2} |\bvec{q}|^2 +\delta \mathcal{E}} \right] \\
	= \underbrace{\int_\mathrm{BZ} \frac{\du^2 \bvec{q}}{(2 \pi)^2} \left[ \frac{\frac{3}{2}|\bvec{q}|^2 - f(\bvec{q})}{\left(\frac{3}{2}|\bvec{q}|^2 + \delta \mathcal{E}\right)(f(q) + \delta \mathcal{E})} \right]}_{=:A} + \underbrace{\int_\mathrm{BZ} \frac{\du^2 \bvec{q}}{(2 \pi)^2} \left[ \frac{1}{\frac{3}{2} |\bvec{q}|^2 +\delta \mathcal{E}} \right]}_{=4(I_1 + I_2)}. \label{eq:i0-split}
\end{multline}
We can verify that the first term in above equation does not have any poles as $\delta \mathcal{E} \to 0$ and hence can be evaluated using \texttt{Mathematica} to yield $A\approx0.0162394$.
In the second term, we can use rotational symmetry to focus on a quarter of the Brillouin zone with $q_x \in [0,\pi]$ and $q_y \in [0,2 \pi/\sqrt{3}]$, which can be divided into two right triangles separated by the line $q_y = 2 q_x /\sqrt{3}$.
These integrals over the two right triangles as domains add up to
\begin{equation}
	I_1 + I_2 = - \frac{1}{(2 \pi)^2} \times \frac{\pi}{ 6} \log \delta \mathcal{E} + 0.0214264 + 0.0190552 + \mathcal{O}(\delta \mathcal{E}),
\end{equation}
so that the asymptotic expansion for the integral $I(0)$ in \eqref{eq:i0-split} may be written as
\begin{equation}
	I(0) = \frac{1}{4 \sqrt{3} \pi} \log \delta \mathcal{E} + C_0 + \mathcal{O}(\delta\mathcal{E}) \quad \text{with} \quad C_0 = -0.196487. \label{eq:asymptot-i}
\end{equation}
One may proceed similarly for the integrals $I_{\beta,\alpha} \equiv I(\bvec{\delta}_{\Au \Bu}^{(\beta)} - \bvec{\delta}_{\Au \Bu}^{(\alpha)})$, where we find that
\begin{equation}
	I_{2,1} = I_{3,2} = I_{3,2} = \frac{1}{4 \sqrt{3} \pi} \log \delta \mathcal{E} +C_1 + \mathcal{O}(\delta \mathcal{E}) \quad \text{with} \quad C_1 = -0.0298208. \label{eq:asymptot-ii}
\end{equation}
We have explicitly verified the asymptotic forms \eqref{eq:asymptot-i} and \eqref{eq:asymptot-ii} by comparing with a numerical evaluation of the required integrals at small but finite $\delta \mathcal{E}$, finding excellent agreement for $\delta \mathcal{E} \lesssim 1$.

\begin{figure}
	\centering
	\includegraphics[width=.5\textwidth]{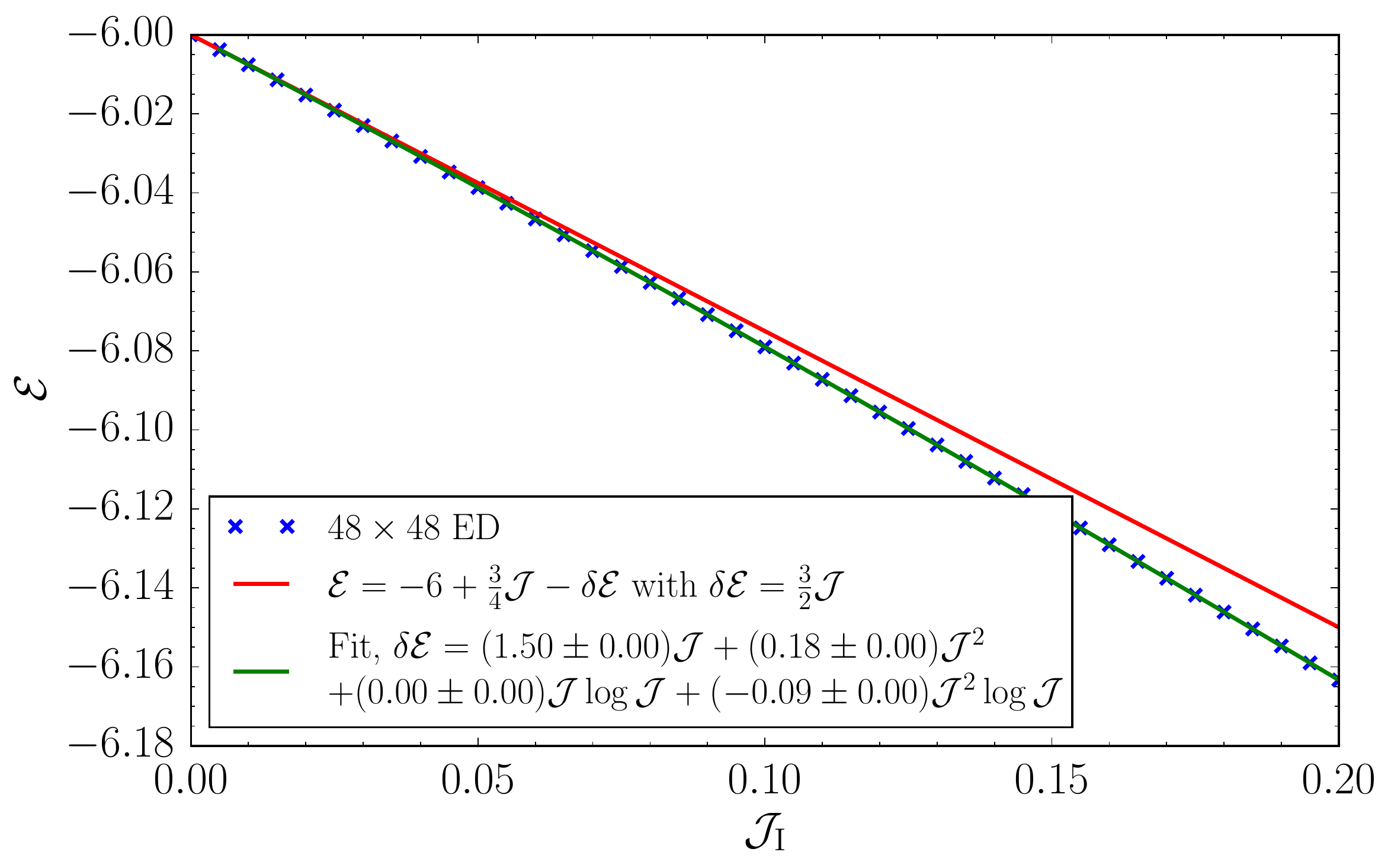}
	\caption{\label{fig:bs-fit}Bound-state energy in $S^z_\mathrm{tot.} = (N-1)/2$ sector obtained via exact diagonalization, analytical form determined from perturbative analysis, and fit of asymptotic form.}
\end{figure}

Eq.~\eqref{eq:bs-matrix-form} may then be rewritten more compactly as
\begin{equation}
	\bigg[\underbrace{\left(\frac{2}{\mathcal{J}_1} + C_0 - C_1 \right)}_{=:\alpha} \mathds{1} + \underbrace{\left( \frac{1}{4 \sqrt{3} \pi} \log\delta\mathcal{E} + C_1 \right)}_{=:\beta} \uvec{u} \uvec{u}^\top \bigg] \begin{pmatrix}
		g_1 \\ g_2 \\ g_3
	\end{pmatrix} = \underbrace{\left(\frac{\sqrt{3}}{4 \pi} \log\delta\mathcal{E} + C_0 + 2C_1 \right)}_{=:\gamma}\uvec{u},
\end{equation}
where $\uvec u = (1,1,1)^\top$. The matrix on the LHS is conveniently inverted using the Sherman-Morrison formula (for $\alpha \beta^{-1} \mathds{1} + \underline{u} \underline{u}^\top)$) and we then find the amplitudes as
\begin{equation} \label{eq:g_i}
	g_i = \frac{\gamma}{3\beta+\alpha} = \frac{\frac{\sqrt{3}}{4 \pi} \log \delta \mathcal{E} + C_0 + 2C_1}{\frac{\sqrt{3}}{4 \pi} \log \delta \mathcal{E} + 2 C_1 + C_0 + \frac{2}{\mathcal{J}_\mathrm{I}}} \quad \text{where} \quad g_i = g(-\bvec{\delta}_{\Au \Bu}^{(i)}).
\end{equation}
We can now plug this into Eq.~(5a) of the main text, and use $\mathcal{E} = - 6 + \frac{3\mathcal{J}_\mathrm{I}}{4} - \delta \mathcal{E}$ on the LHS of Eq.~(5a) of the main text to get
\begin{equation}
	\left( \frac{\sqrt{3}}{ 4 \pi} \log \delta \mathcal{E} + 2 C_1 + C_0 + \frac{2}{\mathcal{J}_\mathrm{I}} \right) \delta \mathcal{E} = 3. \label{eq:bs-energy}
\end{equation}
This is an implicit equation for $\delta \mathcal{E}$ as a function of $\ji$, which can be solved perturbatively in $\ji$. To this end, we first write
\begin{equation}
	\delta \mathcal{E} = \frac{3}{\frac{2}{\mathcal{J}_\mathrm{I}} + 2C_1 + C_0} - \frac{\sqrt{3}}{4 \pi} \frac{\delta \mathcal{E} \log \delta \mathcal{E}}{\frac{2}{\mathcal{J}_\mathrm{I}} + 2C_1 + C_0}.
\end{equation}
Noting that the second term approaches $\delta \mathcal{E} \to 0$ slower that $\delta\mathcal{E}$, we can focus on the first term, yielding
\begin{equation}
	\delta\mathcal{E} \approx \frac{3 \ji}{2} + \mathcal{O}(\ji^2)
\end{equation}
It is expected that the contribution to this asymptotic expansion will appear at order $\ji$ (from expanding the fraction in the first term, and only then we expect some logarithmic corrections $\sim \ji^2 \ln \ji$.

We verify this asymptotic form by extracting the bound-state energy as a function of $\ji$ from exact diagonalization of $\underline{\mathcal{H}}_{\bvec{q}=0}$ on a finite-size cluster of $48 \times 48$ unit cells.
We fit an (unbiased) asymptotic form $\delta \mathcal{E} = A_1 \ji + A_2 \ji \ln \ji + A_3 \ji^2 + A_4 \ji^2 \ln \ji$ with a priori undetermined coefficients $A_1, \dots, A_4$, as shown in Fig.~\ref{fig:bs-fit}, which confirms the leading-order term $A_1 = 3/2$ and the absence of the first logarithmic correction, $A_2 = 0$.

\section{Strong-coupling analysis}

The three $\Au$ sublattice local moments near a $\Bu$ site can be split into 2 $\mathsf{S}=1/2$ representations and a $\mathsf{S}=3/2$ representation.
Writing the Kondo interaction as
\begin{equation}
	\mathcal{H}  = J \vec{\mathsf{S}} \cdot \vec{s}  = \frac{J}{2} \left( S(S+1) - s(s+1) - \mathsf{S}(\mathsf{S}+1) \right),
\end{equation}
where $s=1/2$ is the spin quantum number of the doped electron with $\vec{s} = \frac{1}{2} c^\dagger_\sigma \vec{\tau}_{\sigma\sigma'} c_\sigma$, and the total spin of the three $\Au$-moments and doped electron is $\vec{S} = \vec{s} + \vec{\mathsf{S}}$, we find that the ground-state energy is minimal for $\mathsf{S}=3/2$ and $S = 1$.

\begin{figure}
	\centering
	\includegraphics[width=.3\textwidth]{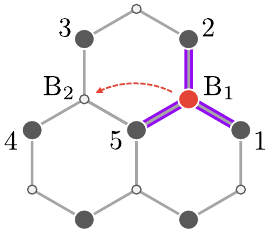}
	\caption{\label{fig:illustrongcoupling}Illustration of hopping process at first-order perturbation theory in the $\mathcal{J}_\mathrm{I} \gg 1$ limit. Purple shading indicates that the B-electron forms a strongly-coupled $S=1$ cluster with its neighboring A-sublattice local moments labelled $1,2,5$. The projected hopping operators maximizes the kinetic energy if the A-moments 3,4 form a $S=1$ representation in the initial state, and the A-moments 1,2 form a $S=1$ representation in the final state of the hopping process.}
\end{figure}

We are interested in degenerate perturbation theory with the perturbation given by the hopping between two $\Bu$-sublattice sites.
The two $\Bu$ sublattice sites $\Bu_1,\Bu_2$ share one common site (see Fig.~\ref{fig:illustrongcoupling} for the labelling of sites).
We can then add the other two spins $\vec{S}_{12} = \vec{S}_1 + \vec{S}_2$ (to write e.g. $\ket{\mathsf{S}^z=3/2} = \ket{S^z_{12}=+1} \ket{S^z_5=1/2}$ if $\vec{\mathsf{S}} = \vec{S}_1 + \vec{S}_2 + \vec{S}_5$), and obtain, using Clebsch-Gordon coefficients,
\begin{subequations}\begin{align}
	\ket{S^z = 1,\Bu_1} &= \sqrt{\frac{3}{4}} \ket{S_{12}^z=+1}\ket{\uparrow_5}\ket{\Downarrow;\Bu_1} - \sqrt{\frac{1}{12}} \ket{S_{12}^z=+1}\ket{\downarrow_5}\ket{\Uparrow;\Bu_1} - \sqrt{\frac{1}{6}} \ket{S_{12}^z=0}\ket{\uparrow_5} \ket{\Uparrow;\Bu_1} \\
	\ket{S^z = 0,\Bu_1} &=  \sqrt{\frac{1}{6}} \Big( \ket{S^z_{12}=+1} \ket{\downarrow_5} \ket{\Downarrow;\Bu_1} - \ket{S^z_{12}=-1} \ket{\uparrow_5} \ket{\Uparrow;\Bu_1} \Big) + \sqrt{\frac{1}{3}} \Big(\ket{S^z_{12} =0} \ket{\uparrow_5} \ket{\Downarrow;\Bu_1} - \ket{S^z_{12} =0} \ket{\downarrow_5} \ket{\Uparrow;\Bu_1} \Big) \\
	\ket{S^z = -1,\Bu_1} &= \sqrt{\frac{1}{12}} \ket{S^z_{12} = -1} \ket{\uparrow_5} \ket{\Downarrow; \Bu_1} + \sqrt{\frac{1}{6}} \ket{S^z_{12} = 0} \ket{\downarrow_5} \ket{\Downarrow;\Bu_1} - \sqrt{\frac{3}{4}} \ket{S_{12}^z = -1} \ket{\downarrow_5} \ket{\Uparrow; \Bu_1}
\end{align}\end{subequations}

We now consider a perturbative process where the doped $B$-electrons hops onto a neighboring site. That is, we are interested in matrix elements (see Fig. \ref{fig:illustrongcoupling} for an illustration)
\begin{equation}
	\bra{S_3,S_4}\braket{S^z_{\Bu_2;125}| \mathcal{H}_t | S^z_{\Bu_1;345}} \ket{S_1,S_2},
\end{equation}
where $\ket{S^z_{\Bu_2;125}}$ denotes that the doped electron is on site $\Bu_1$ and forms a $S_{125}=1$-state with its three neighboring $A$-moments (with $\vec{S}_{\Bu_2;125} = \vec{s}_{\Bu_2} + \vec{S}_1 + \vec{S}_2 + \vec{S}_5$ and similarly for $\vec{S}_{345}$).
If we assume that the dispersion $\mathcal{H}_t$ conserves spin, then its only effect is to move the doped electron from sites $\Bu_1$ to $\Bu_2$ (or vice versa).
We then get ``partial'' matrix elements (the full matrix elements are obtained by applying $\ket{S^z_{12}}$ from the right and $\bra{S^z_{34}}$ from the left) by computing the remaining overlaps,
\begin{subequations}\begin{align}
	\braket{S^z_{\Bu_2;125}=+1| \mathcal{H}_t | S^z_{\Bu_1;345}=+1} &= -t \left( \frac{5}{6} \ket{S^z_{34}=+1}\bra{S^z_{12}=+1} + \frac{1}{6} \ket{S^z_{34}=0}\bra{S^z_{12}=0} \right) \\
	\braket{S^z_{\Bu_2;125}=+1| \mathcal{H}_t | S^z_{\Bu_1;345}=0} &= -t \left( \frac{1}{6} \ket{S^z_{34} = -1} \bra{S^z_{12} = 0} + \frac{2}{3} \ket{S^z_{34} = 0} \bra{S^z_{12}=+1} \right)\\
	\braket{S^z_{\Bu_2;125}=0| \mathcal{H}_t | S^z_{\Bu_1;345}=+1,\Bu_1} &= -t \left( \frac{2}{3} \ket{S^z_{34} = +1} \bra{S^z_{12} = 0} + \frac{1}{6} \ket{S^z_{34}=0}\bra{S^z_{12} =-1} \right) \\
	\braket{S^z_{\Bu_2;125}=0 | \mathcal{H}_t | S^z_{\Bu_1;345}=0} &= -t\left[\frac{1}{6} \big(\ket{S^z_{12}=+1}\bra{S^z_{34}=+1} +  \ket{S^z_{12}=-1}\bra{S^z_{34}=-1} \big) + \frac{2}{3} \ket{S^z_{12}=0} \bra{S^z_{34} = 0} \right] \\
	\braket{S^z_{\Bu_2;125} =+1 | \mathcal{H}_t | S^z_{\Bu_1;345} = -1 } &= -t \left( \frac{1}{2} \ket{S^z_{34} = -1}\bra{S^z_{12}=+1} \right) \\
	\braket{S^z_{\Bu_2;125} =-1 | \mathcal{H}_t | S^z_{\Bu_1;345} = +1 } &= -t \left( \frac{1}{2} \ket{S^z_{34} = +1}\bra{S^z_{12}=-1} \right),
\end{align}\end{subequations}
and all other matrix elements are obtained by time reversal (i.e. $S^z = +1 \to S^z=-1$ and $\Uparrow\to\Downarrow$ etc.), or by hermiticity.
In general, the Hamiltonian projected into the degenerate manifold then has the form
\begin{equation}
	\underline{\mathcal{H}}_t = \begin{pmatrix}
		0 & H_t^{\Bu_2 \leftarrow \Bu_1} \\
		\left( H_t^{\Bu_2 \leftarrow \Bu_1} \right)^\dagger & 0
	\end{pmatrix},
\end{equation}
where we order the basis states as
\begin{multline}
	\Big\{\ket{S^z_{\Bu_1;345} = +1, \Bu_1}\ket{S_{12}^z = +1}, \dots, \ket{S^z_{\Bu_1;345} = +1, \Bu_1}\ket{S_{12}^z = -1}, \ket{S^z_{\Bu_1;345} = 0, \Bu_1}\ket{S_{12}^z = +1}, \dots, \\ 
	\ket{S^z_{\Bu_1;345} = -1, \Bu_1}\ket{S_{12}^z = -1}, \dots, \ket{S^z_{\Bu_2;125} = +1, \Bu_2}\ket{S_{34}^z = +1}, \dots, \ket{S^z_{\Bu_2;125} = -1, \Bu_2}\ket{S_{34}^z = -1} \Big\}
\end{multline}
so that we have
\begin{equation} \label{eq:ht-block}
	H_t^{\Bu_2 \leftarrow \Bu_1} =  -t\left(\begin{array}{ccc|ccc|ccc}
		\frac{5}{6} & 0 & 0  & 0 & 0 & 0  & 0 & 0 & 0 \\ 
		0 & \frac{1}{6} & 0  &\frac{2}{3}& 0 & 0 & 0 & 0 & 0 \\  
		0 & 0 & 0 & 0 & \frac{1}{6} & 0 & \frac{1}{2} & 0 & 0 \\  \hline
		0 & \frac{2}{3} & 0 & \frac{1}{6} & 0 & 0 & 0 &0 &0 \\
		0 & 0 & \frac{1}{6} & 0 & \frac{2}{3} & 0 & \frac{1}{6} & 0 & 0 \\
		0 & 0 & 0 & 0 & 0 &\frac{1}{6} & 0 &\frac{2}{3} & 0 \\ \hline
		0 & 0 & \frac{1}{2} & 0 & \frac{1}{6} & 0 & 0 & 0 & 0 \\
		0 & 0 & 0 & 0 & 0 & \frac{2}{3}  & 0 &\frac{1}{6} & 0 \\
		0 & 0 & 0 & 0 & 0 & 0 & 0 & 0 & \frac{5}{6}	\end{array} \right).
\end{equation}
In principle we can now simply diagonalize $\underline{\mathcal{H}}_t$, but this will also diagonalize the hopping term (i.e. we get some hybridization between $\Bu_1$ and $\Bu_2$), which does not generalize well to the lattice system with multiple neighbors.
Instead, it is convenient to diagonalize the hopping term only in terms of its spin quantum numbers, so that the upper right and lower left blocks of $\mathcal{H}_t$ are diagonal, and we have multiple ``channels'' of hopping, depending on the initial/final configuration of the spins on sites 1,2 and 3,4 -- the channel with the lowest energy (largest in magnitude) will be most important for the ground state.

Diagonalizing \eqref{eq:ht-block}, we get the spectrum with the notation ``$(\text{Eigenvalue},\text{multiplicity})$'' as
\begin{equation}
	\{ \left( \frac{-5 t}{6}, 5 \right), \left(-\frac{t}{3},1\right), \left( \frac{t}{2},3 \right)\},
\end{equation}
where the eigenvectors int the five-fold degenerate subspace are given by
\begin{subequations}\begin{align}
	\ket{+2} &= \ket{S^z_{\Bu_1;345}=+1} \ket{S^z_{12} = +1} \\
	\ket{+1} &= \frac{1}{\sqrt{2}} \big( \ket{S^z_{\Bu_1;345}=+1} \ket{S^z_{12} = 0} + \ket{S^z_{\Bu_1;345}=0} \ket{S^z_{12} = +1} \big) \\
	\ket{0} &= \frac{1}{\sqrt{6}} \ket{S^z_{\Bu_1;345}=+1} \ket{S^z_{12}=-1} + \sqrt{\frac{2}{3}} \ket{S^z_{\Bu_1;345}=0} \ket{S^z_{12}=0} + \frac{1}{\sqrt{6}} \ket{S^z_{\Bu_1;345}=-1} \ket{S^z_{12}=+1} \\
	\ket{-1} &= \frac{1}{\sqrt{2}} \big( \ket{S^z_{\Bu_1;345}=0} \ket{S^z_{12} = -1} + \ket{S^z_{\Bu_1;345}=-1} \ket{S^z_{12} = 0} \big) \\
	\ket{-2} &= \ket{S^z_{\Bu_1;345}=-1} \ket{S^z_{12} = -1},
\end{align}\end{subequations}
and similarly for the $S^z_{\Bu_2;125}$ and $S^z_{34}$.
These states are the five $S_\mathrm{tot} = 2$ states for full cluster (one $c$ electron and 5 local moments) that one gets from adding $S_{\Bu_1;345}=1$ and $S_{12}=1$.
This implies that there is a dispersing cluster (consisting of three local moments and the doped electron) with $S=1$ (which is less that the maximum possible value $S = 3\times (1/2) + 1/2 = 2$, corresponding to either a flipped local moment or a $c_\downarrow$-electron), which maximizes kinetic energy if the remaining spins are in a configuration that maximizes their spin.
This is consistent with the previously found spin-polaron dispersing in a polarized (ferromagnetic) background.

\section{Mean-field theory with local moments}

Here, we treat the Kondo interaction $J_\mathrm{K}^{(\mathrm{I})}$ using a mean-field approximation where we  treat the spins as classical local moments and variationally seek the spin configuration which minimizes the ground-state energy of the full system.
To this end, we consider the Hamiltonian $\mathcal{H} = \mathcal{H}_t + \mathcal{H}_{J_\mathrm{K}^{(\mathrm{I})}}$ where
\begin{align}
	\mathcal{H}_{J_\mathrm{K}^{(\mathrm{I})}} &= J_\mathrm{K}^{(\mathrm{I})} \sum_i \sum_{\delta_{\Au \Bu}} \frac{1}{2} c_{i,\alpha}^\dagger \vec{\sigma}_{\alpha,\beta} c_{i,\beta} \cdot \vec{S}_{i+\delta_{\Au \Bu}} \\
	&= \frac{J_\mathrm{K}^{(\mathrm{I})}}{2} \sum_i \sum_{\delta_{\Au \Bu}} \left( c_{i,\uparrow}^\dagger c_{i,\downarrow} S^-_{i+\delta_{\Au \Bu}} + c_{i,\downarrow}^\dagger c_{i,\uparrow} S^+_{i+\delta_{\Au \Bu}} + \left(c_{i,\uparrow}^\dagger c_{i,\uparrow} - c_{i,\downarrow}^\dagger c_{i,\downarrow} \right) S^z_{i+\delta_{\Au \Bu}}\right)
\end{align}
For concreteness, we resrict ourselves to single-$\bvec{Q}$ order, and by global $\SUtwo$ symmetry, we can take the ordered spins to lie in the plane, such that the classical approximation amounts to writing
\begin{equation}
	S_i^+ = S \eu^{\iu \bvec{Q} \cdot r_i}, \quad S_i^- = S \eu^{-\iu \bvec{Q} \cdot r_i} \quad \text{and} \quad S^z_i = 0.
\end{equation}
Fourier-transforming with $c_i = \frac{1}{\sqrt{N}} \sum_k \eu^{\iu k \cdot r_i} c_k$, we get
\begin{equation}
	\mathcal{H}_{J_\mathrm{K}^{(\mathrm{I})}} = \frac{JS}{2} \sum_{\bvec{k}} \sum_{\delta_{\Au \Bu}} \left(\eu^{-\iu \bvec{Q} \cdot \delta_{\Au \Bu}} c_{\bvec{k}\uparrow}^\dagger c_{\bvec{k}+\bvec{Q}\downarrow} + \eu^{\iu \bvec{Q}\cdot \delta_{\Au \Bu}} c_{\bvec{k}+\bvec{Q}\downarrow}^\dagger c_{k\uparrow} \right).
\end{equation}
With $f_{\Au \Bu}(\bvec{Q}) = \sum_{\delta{\Au \Bu}} \eu^{\iu \bvec{Q}\cdot \bvec{\delta}_{\Au \Bu}}$ we can then write down the full Hamiltonian as 
\begin{equation} \label{eq:hmat-t-xi}
	\mathcal{H} = \mathcal{H}_t +  \mathcal{H}_{J_\mathrm{K}^{(\mathrm{I})}}   = \sum_k \begin{pmatrix}
		c_{\bvec{k} \uparrow}^\dagger & c_{\bvec{k}+\bvec{Q}\downarrow}^\dagger
	\end{pmatrix} \begin{pmatrix}
		\xi(k) & \frac{JS}{2} f_{\Au \Bu}^\ast(\bvec{Q}) \\
		\frac{JS}{2} f_{\Au \Bu}(\bvec{Q}) & \xi(\bvec{k}+\bvec{Q})
	\end{pmatrix} \begin{pmatrix}
		c_{\bvec{k} \uparrow} \\ c_{\bvec{k}+\bvec{Q}\downarrow}
	\end{pmatrix},
\end{equation}
where $\xi(\bvec{k}) = -2 t \Re[f_{\Bu \Bu}(\bvec{k})]$ is the dispersion of the (free) doped electrons on the effective triangular lattice spanned by the $B$ sublattice sites.
We now diagonalize \eqref{eq:hmat-t-xi}, giving the levels
\begin{equation}
	\varepsilon^\pm_{\bvec{Q}}(\bvec{k}) =  \frac{1}{2} \left[\xi(\bvec{k}) + \xi(\bvec{k}+\bvec{Q}) \pm \sqrt{\left(\xi(\bvec{k}) - \xi(\bvec{k}+\bvec{Q}) \right)^2 + (JS)^2 |f_{\Au \Bu}(\bvec{Q})|^2 } \right].
\end{equation}
Working explicitly on a lattice of $L \times L$ unit cells (i.e. $L \times L$ effective triangular lattice sites for the $B$ electrons), giving rise to $2 L^2$ electronic states, we now fill the lowest lying $N_\mathrm{fill}$ levels (with $N_\mathrm{fill}$ even to account for spin degeneracy) to obtain the ground-state energy $E_{\bvec{Q}}$ as a function of the ordering wavevector $\bvec{Q}$.
We find that independent of the chosen filling and for all $\Jki/t \neq 0$, $E(\bvec{Q}=0)$ minimizes the ground state energy, implying ferromagnetic order.

\section{Resummation of RKKY interaction}

One subtlety is that the sum over sites appearing in $\mathcal{H}_\mathrm{RKKY}$ in Eq.~(7) of the main text is not only double-counting each bond (as usual), but the choice of indexing does not allow for a straightfoward identification of $n$-th nearest-neighbor interactions of local moments on the triangular $\Au$ sublattice, because e.g. $\bvec{r}_i = \bvec{r}_j - \bvec{n}_2$ with $\bvec{\delta}_{\Au \Bu} = \bvec{\delta}_{\Au \Bu}' = 0$ index the same interaction as $\bvec{r}_i = \bvec{r}_j$ and $\bvec{\delta}_{\Au \Bu} = -\bvec{n}_2$ and $\bvec{\delta}_{\Au \Bu}'=0$.
So we expand the sum in $\mathcal{H}_\mathrm{RKKY}$ relative to some site $i$ in \texttt{Mathematica} and use pattern matching to reorganize it.
We further note that by the lattice symmetries of the system, we have $J_\mathrm{RKKY}(\pm \bvec{n}_1) = J_\mathrm{RKKY}(\pm \bvec{n}_2) = J_\mathrm{RKKY}(\pm (\bvec{n}_1-\bvec{n}_2)) \equiv J_\mathrm{RKKY}^{(1)}$ and simlarly for $J_\mathrm{RKKY}^{(2)}$, and we define the shorthand $J_\mathrm{RKKY}^{(0)} = J_\mathrm{RKKY}(0)$, so that we can rewrite
\begin{multline}
	\sum_{i,j} \sum_{\delta_{\Bu_2},\delta_{\Au \Bu}'} J_\mathrm{RKKY}(\bvec{r}_i - \bvec{r}_j) \vec S_{i+\delta_{\Au \Bu} } \cdot \vec{S}_{j+\delta_{\Au \Bu}'} = \frac{1}{2}\sum_{\langle ij \rangle} \left( J_\mathrm{RKKY}^{(0)} + 5 J_\mathrm{RKKY}^{(1)} + 2 J_\mathrm{RKKY}^{(2)}  \right) \vec{S}_i \cdot \vec{S}_j \\
	+ \frac{1}{2}\sum_{\langle \langle ij \rangle \rangle} \left( 2 J_\mathrm{RKKY}^{(1)} +3 J_\mathrm{RKKY}^{(2)} \right) \vec{S}_i \cdot \vec{S}_j + \dots + \mathrm{const.},
\end{multline}
where $\langle \dots \rangle_\mathrm{A}$ and $\langle\langle \dots \rangle \rangle_\mathrm{A}$ refers to nearest and second-nearest neighboring sites on the $\Au$ sublattice.

\section{Two-band model in a magnetic field}

From the kinetic term, we get the matrix elements
\begin{align}
	\braket{c_i| \mathcal{H}_t | c_{i+\delta_{\Bu \Bu}}} &= -t \\
	\braket{d_i| \mathcal{H}_t | d_{i+\delta_{\Bu \Bu}}} &= \frac{-t}{|\beta|^2} \left( 1 + \sum_l g^\ast(\bvec{r}_i - \bvec{r}_l) g(\bvec{r}_i + \bvec{\delta}_{\Bu \Bu} - \bvec{r}_l) \right), \label{eq:dhop-realspace}
\end{align}
where $\delta_{\Bu \Bu}$ are the three lattice vectors for nearest neighbors on the $\Bu $ sublattice, and $\beta = \sqrt{N}/\mathcal{N}$.
We can further consider the Kondo term $\mathcal{H}_{J_\mathrm{K}^{(\mathrm{I})}}$, which just gives onsite terms in the two bands,
\begin{align}
	\braket{c_i| \mathcal{H}_{J_\mathrm{K}^{(\mathrm{I})}} | c_i} &= \frac{3 J_\mathrm{K}^{(\mathrm{I})}}{4} \\
	\braket{d_i| \mathcal{H}_{J_\mathrm{K}^{(\mathrm{I})}} | d_i} &= \frac{J_\mathrm{K}^{(\mathrm{I})}}{|\beta|^2} \left( -\frac{3}{4} + \Re \left[\sum_{\delta_{\Au \Bu}} g(-\delta_{\Au \Bu}) \right] + \frac{3}{4} \sum_l |g(\bvec{r}_i -\bvec{r}_l)|^2 - \frac{1}{2} \sum_{\delta_{\Au \Bu}} |g(-\delta_{\Au \Bu})|^2  \right).
\end{align}
Applying a magnetic field (along $\hat{z}$ wlog) amounts to including the matrix elements of the Hamiltonian
\begin{equation}
	\mathcal{H}_h = - h \left( \sum_{i\in A} S^z_i + \sum_{i \in B} \frac{1}{2} c_{i,\sigma}^\dagger \vec{\tau}_{\sigma,\sigma'} c_{i,\sigma'} \right) = - h \sum_i \left( \frac{n_{i,\uparrow} - n_{i,\downarrow}}{2} + \frac{1}{2} - n^{(b)}_i \right),
\end{equation}
and we can pull out the global constant $E_\mathrm{const.} = - h N/2$ which just corresponds to the energy of $N$ local moments aligned along the field.
Then the Hamiltonian has matrix elements
\begin{align}
	\braket{c_i| \mathcal{H}_{J_\mathrm{K}^{(\mathrm{I})}} | c_i} &= -\frac{h}{2} \\
	\braket{d_i| \mathcal{H}_{J_\mathrm{K}^{(\mathrm{I})}} | d_i} &= \frac{h}{2}
\end{align}
This tight-binding Hamiltonian is readily diagonalized by going into momentum space, where matrix elements are given by $\braket{c_{\bvec{k}}|\mathcal{H}|c_{\bvec{q}}} = \frac{1}{N} \sum_{i,j} \braket{c_i | \mathcal{H} | c_j} \eu^{\iu \left(-\bvec{k} \cdot \bvec{r}_i + \bvec{q} \cdot \bvec{r}_j\right)}$
and similar for the $d$-particles.
We find the dispersion for the $c_\uparrow$ electrons to be given by
\begin{equation}
	\varepsilon_{c_\uparrow}(\bvec{k}) = -2 t \Re\left[ f_{\Bu \Bu}(\bvec{k}) \right] + \frac{3 J_\mathrm{K}^{(\mathrm{I})}}{4} - \frac{h(N+1)}{2}.
\end{equation}
From \eqref{eq:dhop-realspace}, we get
\begin{equation}
	\braket{d_{\bvec{k}} | \mathcal{H}_t | d_{\bvec{q}}} = - \frac{t}{|\beta|^2} \delta_{\bvec{k},\bvec{q}} \left(2 \Re\left[ f_{\Bu \Bu}(\bvec{k}) \right] + \frac{1}{N} \sum_{\bvec{p}} 2 \Re[f_{\Bu \Bu}(\bvec{p}+\bvec{q})] \times |\tilde{g}(\bvec{p})|^2  \right),
\end{equation}
where $\tilde{g}(\bvec{p})$ denotes Fourier components of the bound-state wavefunction that can be obtained numerically, or by using an analytical solution of Eqs.~(5a)~and~(5b) (note that this limits us to working only perturbatively in $\ji$), and we further have the matrix elements
\begin{equation}
	\braket{d_{\bvec{k}} | \mathcal{H}_{J_\mathrm{K}^{(\mathrm{I})}} | d_{\bvec{q}}} = \delta_{\bvec{k},\bvec{q}} \frac{J_\mathrm{K}^{(\mathrm{I})}}{|\beta|^2} \left( - \frac{3}{4} + \Re\left[ \sum_{\delta_{\Au \Bu}} g(-\bvec{\delta}_{\Au \Bu}) \right] + \frac{3}{4 N} \sum_p | \tilde{g}(\bvec{p})|^2 - \frac{1}{2} \sum_{\delta_{\Au \Bu}} | g(-\bvec{\delta}_{\Au \Bu})|^2 \right)
\end{equation}


\bibliography{TMD_polaron_bib}
\bibliographystyle{apsrev4-2}

\clearpage
\onecolumngrid
\appendix